\documentclass[twocolumn]{aastex631}

\def\arcmin{^{\prime}}
\def\farcs{\hbox{$.\!\!^{\prime\prime}$}}

\def\farc{\hbox{$\ \!\!^{\prime\prime}$}}
\def\h{$^{\mathrm{h}}$}
\def\m{$^{\mathrm{m}}$}
\def\s{$^{\mathrm{s}}$}
\def\tlvco{$^{12}$CO}
\def\thrco{$^{13}$CO}
\def\C18O{C$^{18}$O}

\begin{document}

\title{Early Planet Formation in Embedded Disks (eDisk) IX: High-resolution ALMA Observations of the Class 0 Protostar R CrA IRS5N and its surrounding}

\correspondingauthor{Rajeeb Sharma}
\email{rajeeb.sharma@nbi.ku.dk}

\author[0000-0002-0549-544X]{Rajeeb Sharma}, 
\affiliation{Niels Bohr Institute, University of Copenhagen, \O ster Voldgade 5--7, 1350, Copenhagen K, Denmark}

\author[0000-0001-9133-8047]{Jes K. J\o rgensen}
\affiliation{Niels Bohr Institute, University of Copenhagen, \O ster Voldgade 5--7, 1350, Copenhagen K, Denmark}

\author[0000-0001-5782-915X]{Sacha Gavino}
\affiliation{Niels Bohr Institute, University of Copenhagen, \O ster Voldgade 5--7, 1350, Copenhagen K, Denmark}

\author[0000-0003-0998-5064]{Nagayoshi Ohashi}
\affiliation{Academia Sinica Institute of Astronomy \& Astrophysics \\
11F of Astronomy-Mathematics Building, AS/NTU, No.1, Sec. 4, Roosevelt Rd \\
Taipei 10617, Taiwan, R.O.C.}

\author[0000-0002-6195-0152]{John J. Tobin}
\affil{National Radio Astronomy Observatory, 520 Edgemont Rd., Charlottesville, VA 22903 USA} 

\author[0000-0001-7233-4171]{Zhe-Yu Daniel Lin}
\affiliation{University of Virginia, 530 McCormick Rd., Charlottesville, Virginia 22903, USA}

\author[0000-0002-7402-6487]{Zhi-Yun Li}
\affiliation{University of Virginia, 530 McCormick Rd., Charlottesville, Virginia 22903, USA}

\author[0000-0003-0845-128X]{Shigehisa Takakuwa}
\affiliation{Department of Physics and Astronomy, Graduate School of Science and Engineering, Kagoshima University, 1-21-35 Korimoto, Kagoshima,Kagoshima 890-0065, Japan}
\affiliation{Academia Sinica Institute of Astronomy \& Astrophysics \\
11F of Astronomy-Mathematics Building, AS/NTU, No.1, Sec. 4, Roosevelt Rd \\
Taipei 10617, Taiwan, R.O.C.}

\author[0000-0002-3179-6334]{Chang Won Lee}
\affiliation{Korea Astronomy and Space Science Institute, 776 Daedeok-daero Yuseong-gu, Daejeon 34055, Republic of Korea}
\affiliation{University of Science and Technology, 217 Gajeong-ro Yuseong-gu, Daejeon 34113, Republic of Korea}

\author[0000-0003-4361-5577]{Jinshi Sai (Insa Choi)}
\affiliation{Academia Sinica Institute of Astronomy \& Astrophysics \\
11F of Astronomy-Mathematics Building, AS/NTU, No.1, Sec. 4, Roosevelt Rd \\
Taipei 10617, Taiwan, R.O.C.}

\author[0000-0003-4022-4132]{Woojin Kwon}
\affiliation{Department of Earth Science Education, Seoul National University, 1 Gwanak-ro, Gwanak-gu, Seoul 08826, Republic of Korea}
\affiliation{SNU Astronomy Research Center, Seoul National University, 1 Gwanak-ro, Gwanak-gu, Seoul 08826, Republic of Korea}

\author[0000-0003-4518-407X]{Itziar de Gregorio-Monsalvo}
\affiliation{European Southern Observatory, Alonso de Cordova 3107, Casilla 19, Vitacura, Santiago, Chile}

\author[0000-0001-6267-2820]{Alejandro Santamaría-Miranda}
\affiliation{European Southern Observatory, Alonso de Cordova 3107, Casilla 19, Vitacura, Santiago, Chile}

\author[0000-0003-1412-893X]{Hsi-Wei Yen}
\affiliation{Academia Sinica Institute of Astronomy \& Astrophysics \\
11F of Astronomy-Mathematics Building, AS/NTU, No.1, Sec. 4, Roosevelt Rd \\
Taipei 10617, Taiwan, R.O.C.}

\author[0000-0003-3283-6884]{Yuri Aikawa}
\affiliation{Department of Astronomy, Graduate School of Science, The University of Tokyo, 7-3-1 Hongo, Bunkyo-ku, Tokyo 113-0033, Japan}

\author[0000-0002-8238-7709]{Yusuke Aso}
\affiliation{Korea Astronomy and Space Science Institute, 776 Daedeok-daero Yuseong-gu, Daejeon 34055, Republic of Korea}

\author[0000-0001-5522-486X]{Shih-Ping Lai}
\affiliation{Academia Sinica Institute of Astronomy \& Astrophysics \\
11F of Astronomy-Mathematics Building, AS/NTU, No.1, Sec. 4, Roosevelt Rd \\
Taipei 10617, Taiwan, R.O.C.}
\affiliation{Institute of Astronomy, National Tsing Hua University, No. 101, Section 2, Kuang-Fu Road, Hsinchu 30013, Taiwan}
\affiliation{Center for Informatics and Computation in Astronomy, National Tsing Hua University, No. 101, Section 2, Kuang-Fu Road, Hsinchu 30013, Taiwan}
\affiliation{Department of Physics, National Tsing Hua University, No. 101, Section 2, Kuang-Fu Road, Hsinchu 30013, Taiwan}

\author[0000-0003-3119-2087]{Jeong-Eun Lee}
\affiliation{Department of Physics and Astronomy, Seoul National University, 1 Gwanak-ro, Gwanak-gu, Seoul 08826, Korea}

\author[0000-0002-4540-6587]{Leslie W. Looney}
\affiliation{Department of Astronomy, University of Illinois, 1002 West Green St, Urbana, IL 61801, USA}

\author[0000-0002-4372-5509]{Nguyen Thi Phuong}
\affiliation{Korea Astronomy and Space Science Institute, 776 Daedeok-daero Yuseong-gu, Daejeon 34055, Republic of Korea}
\affiliation{Department of Astrophysics, Vietnam National Space Center, Vietnam Academy of Science and Techonology, 18 Hoang Quoc Viet, Cau Giay, Hanoi, Vietnam}

\author[0000-0003-0334-1583]{Travis J.\ Thieme}
\affiliation{Institute of Astronomy, National Tsing Hua University, No. 101, Section 2, Kuang-Fu Road, Hsinchu 30013, Taiwan}
\affiliation{Center for Informatics and Computation in Astronomy, National Tsing Hua University, No. 101, Section 2, Kuang-Fu Road, Hsinchu 30013, Taiwan}
\affiliation{Department of Physics, National Tsing Hua University, No. 101, Section 2, Kuang-Fu Road, Hsinchu 30013, Taiwan}

\author[0000-0001-5058-695X]{Jonathan P. Williams}
\affiliation{Institute for Astronomy, University of Hawai‘i at Mānoa, 2680 Woodlawn Dr., Honolulu, HI 96822, USA}





\begin{abstract}
We present high-resolution, high-sensitivity observations of the Class 0 protostar RCrA IRS5N as part of the Atacama Large Milimeter/submilimeter Array (ALMA) large program Early Planet Formation in Embedded Disks (eDisk). The 1.3 mm continuum emission reveals a flattened continuum structure around IRS5N, consistent with a protostellar disk in the early phases of evolution. The continuum emission appears smooth and shows no substructures. However, a brightness asymmetry is observed along the minor axis of the disk, suggesting the disk is optically and geometrically thick. We estimate the disk mass to be between 0.007 and 0.02 M$_{\odot}$.  Furthermore, molecular emission has been detected from various species, including \C18O~(2--1), \tlvco~(2--1), \thrco~(2--1), and H$_2$CO~(3$_{0,3}-2_{0,2}$, 3$_{2,1}-2_{2,0}$, and 3$_{2,2}-2_{2,1}$). 
By conducting a position-velocity analysis of the \C18O~(2--1) emission, we find that the disk of IRS5N exhibits characteristics consistent with Keplerian rotation around a central protostar with a mass of approximately 0.3 M$_{\odot}$. Additionally, we observe dust continuum emission from the nearby binary source, IRS5a/b. The emission in \tlvco~toward IRS5a/b seems to emanate from IRS5b and flow into IRS5a, suggesting material transport between their mutual orbits. The lack of a detected outflow and large-scale negatives in \tlvco~observed toward IRS5N suggests that much of the flux from IRS5N is being resolved out. Due to this substantial surrounding envelope, the central IRS5N protostar is expected to be significantly more massive in the future. 

\end{abstract}



\section{Introduction} \label{sec:intro}

Protostellar disks form as an outcome of the conservation of angular momentum during the gravitational collapse of the dust and gas in the envelope surrounding young stars \citep[e.g.,][]{Terebey_1984,Mckee_2007}. These disks not only regulate the mass accreted onto the protostar but also provide the necessary ingredients for planet formation \citep{Testi_2014}. Recent Atacama Large Milimeter/submilimeter Array (ALMA) observations with high spatial resolution have discovered that substructures such as gaps and rings are common in the dust emission of Class II young stellar object disks \citep{Alma_2015,Andrews_2018,Cieza_2021}. While these structures can be attributed to features such as snowlines and dust traps \citep{Zhang_2015,Gonzalez_2017}, they are largely thought to be indications of embedded planets \citep{Dong_2015,Zhang_2018}. The direct imaging of possible protoplanets in the gap of the continuum emission of the protostar PDS 70 further supports this idea \citep{Keppler_2018,Isella_2019,Benisty_2021}. 

Recent studies have shown that the mass reservoir of Class II disks is generally insufficient to form giant planets \citep{Tychoniec_2020}. This suggests that planet formation is already well underway by the time a protostar reaches the Class II (T~Tauri) phase. Interferometric observations over the last decade have shown that protostellar disks can be found in younger Class 0/I protostars \citep[e.g.,][]{Tobin_2012,Brinch_2013,Ohashi_2014,Sheehan_2017,Sharma_2020,Tobin_2020}. These disks are generally found to be larger and possibly more turbulent compared to disks around more evolved sources \citep{Sheehan_2017,Tychoniec_2020}. Furthermore, evidence of substructures has been observed in a handful of embedded Class I sources \citep[e.g.,][]{Sheehan_2017,Segura_2020,Sheehan_2020}. These results, combined with the ubiquity of substructures in Class II disks, suggest that planet formation likely begins earlier during the Class 0/I phase when the disk is still embedded in its natal envelope. 

To constrain how and when substructures form in young ($\lesssim$1 Myr old) protostellar disks and ultimately understand their nature, a sample of 19 nearby Class 0/I protostellar systems have been studied with ALMA as part of the Large Program Early Planet Formation in Embedded Disks \citep[eDisk;][]{Ohashi_2023}. One of these deeply embedded protostars located in the R Coronae Australis (R CrA) region, the most active star formation region in the Corona Australis molecular cloud, is the Class 0 source RCrA IRS5N (hereafter IRS5N; \citealt{Harju_1993,Chini_2003}). IRS5N \citep[also referred to as CrA-20;][]{Peterson_2011} is part of a group of a dozen deeply embedded young stellar objects (YSOs) in a cluster dubbed the Coronet in the R CrA region \citep{Taylor_1984}. Traditionally, the cluster is estimated to be at a distance of $\sim$ 130 pc. However, from the recent \textit{Gaia} DR2 parallax measurements, the distance to the cluster has been updated to 147 $\pm$ 5 pc \citep{Zucker20}, which we have adopted for this paper. This value is consistent with the distance of 149.4 $\pm$ 0.4 pc measured recently by \citet{Galli20}.

The Coronet has been extensively observed from X-rays to radio wavelengths \cite[e.g.,][see also review by \citealt{Neuhauser_2008}]{Peterson_2011,Lindberg_2014,Sandell_2021}. Based on {\it Spitzer} photometry of the Coronet, IRS5N was first classified as a Class I source \citep{Peterson_2011}, which was later updated to Class 0 with the addition of {\it Herschel} and JCMT/SCUBA data \citep{Lindberg_2014}. From a recent reanalysis of the spectral energy distribution (SED) of IRS5N utilizing the most recent photometry and the updated Gaia distance above, we find its bolometric temperature ($T_{\mathrm{bol}}$) = 59 K and its bolometric luminosity ($L_{\mathrm{bol}}$) = 1.40 $L_{\odot}$ \citep{Ohashi_2023}. 
Up to now, the highest angular resolution observations of IRS5N at submillimeter wavelengths so far were from the Submillimeter Array (SMA) in the compact configuration at a resolution of 4\farcs6 $\times$ 2\farcs6 \citep{Peterson_2011}. In this paper, part of the series of first-look papers from eDisk, we present the first high-angular resolution ($\sim$0\farcs05), high-sensitivity continuum and spectral line observations toward IRS5N using ALMA. The field-of-view of our ALMA observations of IRS5N also captures the nearby binary protostar, IRS5 a and b. IRS5 (also known as R CrA 19; \citet{Peterson_2011}) was first reported in \citet{Taylor_1984} and later found to be a binary \citep{Chen_1993,Nisini_2005}.

The paper is structured as follows: The observations and the data reduction process are described in Sect.~\ref{sec:data}. The empirical results from the observations of the disk continuum and the molecular line emission are presented in Sect.~\ref{sec:results}. The implications of the results are discussed in Sect.~\ref{sec:discussion} and the conclusions are presented in Sect.~\ref{sec:conclusion}.

\section{Observations and Data Reduction}\label{sec:data}

IRS5N was observed as part of the eDisk ALMA large program (2019.1.00261.L, PI: N. Ohashi) in Band 6 at 1.3 mm wavelength. The short-baseline observations were conducted on 2021 May 4 and on 2021 May 15 for a total on-source time of $\sim$76 minutes. The long-baseline observations were made between 2021 August 18 and October 2 for a total on-source time of $\sim$256 minutes. The shortest and the longest projected baselines were 15 m and 11,615 m, respectively. Along with the continuum, molecular line emissions from \tlvco, \thrco, \C18O, SO, SiO, DCN, $c$-C$_3$H$_2$, H$_2$CO, CH$_3$OH, and DCN were also targeted. A detailed description of the observations 
along with the spectral setup, correlator setup, and calibration is provided in \citet{Ohashi_2023}.

The ALMA pipeline-calibrated long- and short-baseline data were further reduced and imaged using the Common Astronomy Software Application (CASA) 6.2.1 \citep{Mcmullin_2007}. The source position was estimated by calculating the continuum peak position for each execution block and aligned to a single phase center when calculating the scaling between the execution blocks. The self-calibration was carried out using the native phase centers of the observations. The short-baseline data were initially self-calibrated with six rounds of phase-only calibration followed by three rounds of phase and amplitude calibration. Then, the long-baseline data were combined with the self-calibrated short-baseline data, and four more rounds of phase-only calibration were performed on the combined data.
The solutions of the continuum self-calibration are applied to the spectral line data as well. 

The final continuum images were created with a range of robust parameters from -2.0 to 2.0. We adopt the robust value of 0.5 for the continuum image in this paper, providing a balance between sensitivity and resolution. This resulted in a synthesized beam of $0\farcs052 \times 0\farcs035$ and an rms noise of 16 $\mu$Jy beam$^{-1}$. The spectral line images are created with robust parameters of 0.5 and 2.0 with {\it uvtaper} = 2000 k$\lambda$. We adopt robust 0.5 for most of the spectral lines except for the \thrco~and H$_2$CO lines, where we adopt robust 2.0 to increase the signal-to-noise ratio. The details of the continuum observations and the detected spectral lines are summarized in Table~\ref{tab:observation}.

\begin{deluxetable*}{cccccccc}
\tablecaption{Overview of the continuum and the detected molecular lines \label{tab:observation}}
\tabletypesize{\small}
\tablehead{
\colhead{Continuum/Molecules} & \colhead{Transition} & \colhead{robust} & \colhead{Frequency} & \colhead{Beam} & \colhead{P.A.} & \colhead{$\Delta v$} & \colhead{RMS}\\
\colhead{} & \colhead{} & \colhead{} & \colhead{(GHz)} & \colhead{($\farc$)} & \colhead{($^{\circ}$)} & \colhead{(km s$^{-1}$)} & \colhead{(mJy beam$^{-1}$)}
}
\startdata
Continuum & -- & 0.5 & 225.000000 & 0.05 $\times$ 0.03 & 60 & -- & 0.016  \\
\C18O & $J = $ 2--1 & 0.5 & 219.560354 & 0.11 $\times$ 0.08 & 83.8 & 0.167 & 1.636  \\
\tlvco & $J = $ 2--1 & 0.5 & 230.538000 & 0.11 $\times$ 0.08 & 85.9 & 0.635 & 0.987  \\
\thrco & $J = $ 2--1 & 2.0 & 220.398684 & 0.15 $\times$ 0.11 & -87.3 & 0.167 & 2.104  \\
H$_{2}$CO & $J_{K_a, K_c} = $ 3$_{0,3}-2_{0,2}$ & 2.0 & 218.222192 & 0.14 $\times$ 0.11 & -86.6 & 1.34 & 0.499  \\
H$_{2}$CO & $J_{K_a, K_c} = $ 3$_{2,1}-2_{2,0}$ & 2.0 & 218.760066 & 0.15 $\times$ 0.11 & -86.8 & 0.167 & 1.471  \\
H$_{2}$CO & $J_{K_a, K_c} = $ 3$_{2,2}-2_{2,1}$ & 2.0 & 218.475632 & 0.17 $\times$ 0.13 & -86.6 & 1.34 & 0.529  \\
\enddata
\end{deluxetable*}


\section{Results}\label{sec:results}

\subsection{Dust continuum emission}

\begin{figure*}[ht!]
  \centering
  \includegraphics[trim={0 2cm 0 2cm},clip,width=1.0\linewidth]{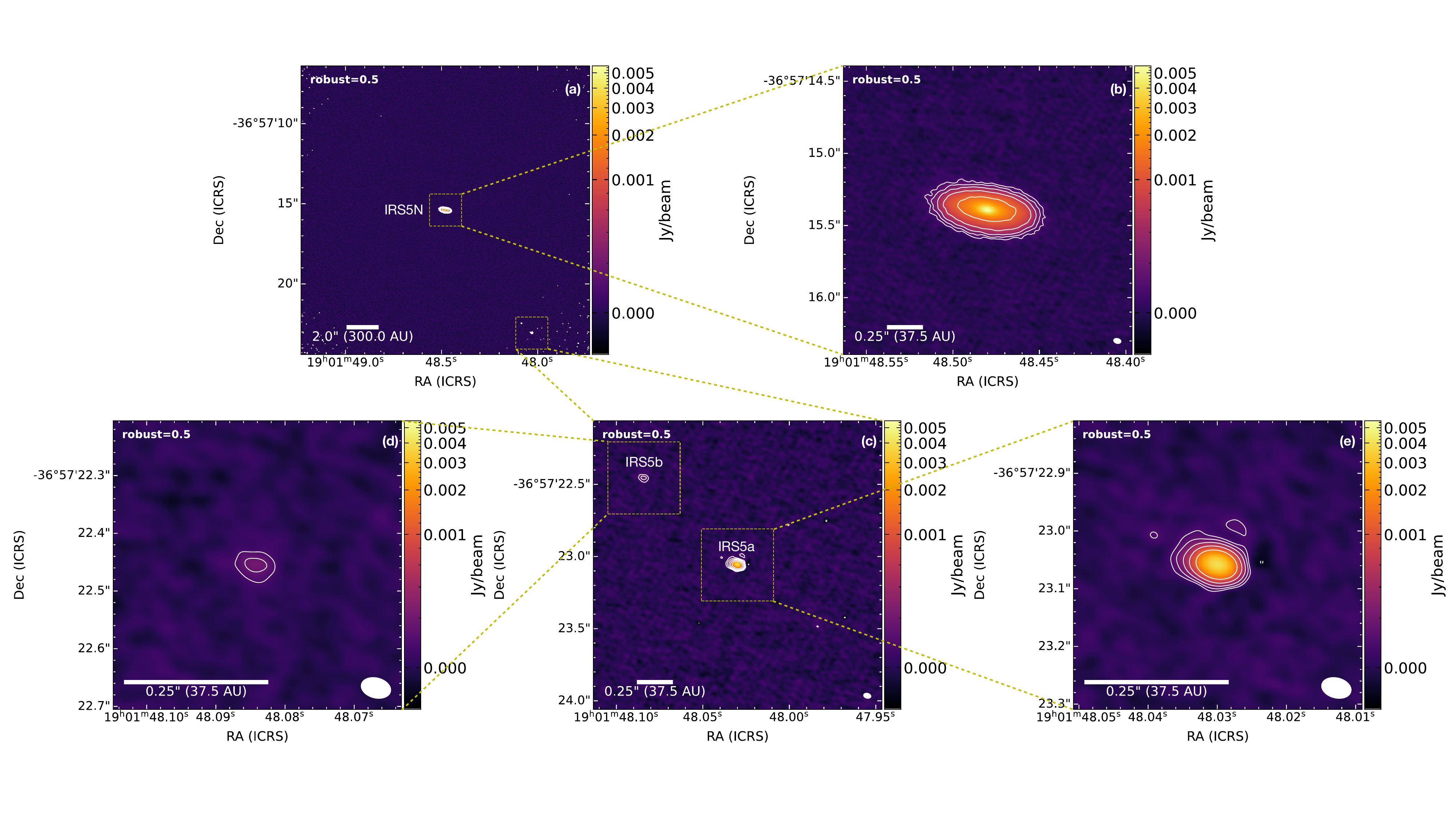}
  \caption{(a) 1.3 mm continuum images of R CrA IRS5N and IRS5 sources with a robust paramater of 0.5. (b) Zoomed-in view of the IRS5N disk. (c) Zoomed-in view of IRS5a and its companion IRS5b. (d) and (e) show further zoomed-in views of the individual sources of the IRS5 binary. The contour levels are 5$\sigma$, 10$\sigma$, 20$\sigma$, 40$\sigma$, and 80$\sigma$ with $\sigma = 0.016$ mJy beam$^{-1}$. The synthesized beam is shown in white in the bottom right corner with a beam size of $0\farcs052 \times 0\farcs035$ and a position angle of 75.4$^{\circ}$. The color stretch used for the images is arcsinh to cover the dynamic range between the sources. \label{fig:cont_irs5n}}
\end{figure*}

Figure~\ref{fig:cont_irs5n} shows the continuum images from the ALMA data at 1.3 mm. Figure~\ref{fig:cont_irs5n}(a) displays the large-scale view of the continuum emission from the region, and the remaining panels show the zoom-in of the IRS5N and the IRS5 protostars.

Figure~\ref{fig:cont_irs5n}(b) shows the zoomed-in view of the IRS5N continuum image. The image shows a well-resolved flattened dust structure, which likely traces the disk surrounding the central protostar. The brightest emission of the disk is concentrated at its geometrical center with a peak intensity of 5.53 mJy beam$^{-1}$ as measured from the emission map corresponding to a brightness temperature of $\sim$94 K, calculated with the full Planck function. The brightness temperature of 94 K is  relatively high for a protostar with $L_{bol}$ = 1.4 $L_\odot$ and deviates from the traditional assumptions of protostars generally derived from Class II disks \citep{Kusaka_1970,Chiang_1997,Huang_2018}. One likely explanation for this high-brightness temperature is that IRS5N experiences self-heating through accretion luminosity, which has also been seen on other eDisk sources and further explored in \citet{Takakuwa_2023}. The total integrated flux density of IRS5N is 101 mJy, measured by integrating pixels where intensity is above 3$\sigma$. The geometrical peak position of IRS5N is $19^\mathrm{h}01^\mathrm{m}48\fs48$, $-36^\circ 57\arcmin 15\farcs 39$. The full width at half maximum (FWHM) of IRS5N is estimated to be $\sim$62 au from the Gaussian fit model of the continuum emission. The deconvolved size enables us to estimate the inclination, $i$, of the IRS5N disk to be $\sim$65$^{\circ}$ calculated from $i = \arccos(\theta_{\rm min}/\theta_{\rm maj})$, where $\theta_{\rm min}$ and $\theta_{\rm max}$ are the FWHM of the minor and major axes respectively.

Figure~\ref{fig:cont_irs5n}(c) shows the zoomed-in view of the binary source IRS5, with panels (d) and (e) showing the zoom-in of IRS5b and IRS5a, respectively. 
\citet{Nisini_2005} first reported a separation of $\sim0\farcs6$ between the two components based on pre-images with a relatively coarse pixel size of 0.14 arcsec/pixel. The pre-images were taken as part of preparations for spectroscopic observations using the ISAAC instrument of the Very Large Telescope (VLT). Our current high-resolution ALMA observations reveal that IRS5a and IRS5b have a projected separation of $\sim$0$\farcs$9 ($\sim$132 au at a distance of 147 pc). This difference between the previous and the new separation may be due to a combination of the proper motions of the sources, and the confusion from the scattered light in the infrared observations. The peak position of IRS5a as measured with Gaussian fitting is $19^\mathrm{h}01^\mathrm{m}48\fs030$, $-36^\circ 57\arcmin 23\farcs 06$. We adopt this position as the coordinate of IRS5a. IRS5a is peaked at the center with a peak intensity of 3.87 mJy beam$^{-1}$ or $\sim$62 K and a flux density of 4.85 mJy. The secondary source, IRS5b, is much smaller and fainter than IRS5a. The peak position of IRS5b as measured with Gaussian fitting is $19^\mathrm{h}01^\mathrm{m}48\fs084$, $-36^\circ 57\arcmin 22\farcs 46$. It has a peak intensity of $\sim$0.20 mJy beam$^{-1}$ or $\sim$3 K and a flux density of 0.26 mJy. The flux density of IRS5 was also measured by integrating above the 3$\sigma$ level over a region surrounding the individual continuum sources. From our observations, IRS5a is marginally resolved whereas IRS5b is not resolved.

\subsection{Disk and envelope masses}
The dust continuum emission with  ALMA can be used to estimate the mass of the total disk structure surrounding the sources. Assuming optically thin emission, well-mixed gas and dust, and isothermal dust emission, the dust mass can be derived from

\begin{equation}\label{eq:disk_mass}
\centering
M_\mathrm{dust} =  \frac{D^2 F_\lambda} {\kappa_{\lambda}B_{\lambda}(T_\mathrm{dust})},
\end{equation}	
where $D$ is the distance to the source ($\sim$147 pc) and $T_{\mathrm{dust}}$ is the temperature of the disk. $F{_\lambda}$, $\kappa_{\lambda}$, and $B_\lambda$ are the flux density of the disk, dust opacity, and the Planck function at the wavelength $\lambda$, respectively. Typically, for Class II disks, $T_{\mathrm{dust}}$ is often taken to be a fixed temperature of 20 K independent of the total luminosity \citep[e.g.,][]{Andrews_2005,Ansdell_2016}. However, for younger, more embedded Class 0/I disks, \citet{Tobin_2020} found through radiative transfer modeling that the dust temperature scales as 

\begin{equation}\label{eq:disk_temp}
\centering
T_{\mathrm{dust}} \approx  \mathrm{43 K} \left(\frac{L_{bol}}{1 L_{\odot}}\right)^{0.25},
\end{equation}
For IRS5N with a bolometric luminosity of 1.40 $L_{\odot}$ Equation (\ref{eq:disk_temp}) yields $T_{\mathrm{dust}}$ = 47 K.

We estimate the disk masses using both dust temperatures. We adopt $\kappa_{\mathrm{1.3mm}}$ = 2.30 cm$^{2}$ g$^{-1}$ from dust opacity models of \citet{Beckwith90} and assume a canonical gas-to-dust ratio of 100:1 to calculate disk masses using Equation~(\ref{eq:disk_mass}). The resulting total disk mass for IRS5N is 0.019 $M_{\odot}$ for a dust temperature of 20 K and 6.65 $\times$ 10$^{-3}$ $M_{\odot}$ for a dust temperature of 47 K. The scaled dust temperature of IRS5a is similar to that of IRS5N, as \citet{Lindberg_2014} found $L_{\mathrm{bol}}$ of IRS5a to be 1.7 $L_\odot$. 
Disk masses are also derived for the binary, IRS5. The estimated disk masses for all the continuum sources are presented in Table \ref{tab:disk_mass}. It is important to note that the disk masses calculated using Equation~(\ref{eq:disk_mass}) represent lower limits, as the continuum emission is most likely optically thick (see Section~\ref{sec:discussion}).

\begin{deluxetable}{cccc}
\tablecaption{Estimated disk masses of continuum emission sources. \label{tab:disk_mass}}
\tablehead{
\colhead{Source} & \colhead{Flux Density} & \multicolumn{2}{c}{Gas+Dust Mass} \\
\colhead{} & \colhead{} & \colhead{20 K} & \colhead{47 K} \\
\colhead{} & \colhead{(mJy)} & \colhead{($M_{\odot}$)} & \colhead{($M_{\odot}$)}
}
\startdata
IRS5N & 100.65 & 0.019 & 6.65 $\times$ 10$^{-3}$ \\
IRS5a & 4.85 & 8.92 $\times$ 10$^{-4}$ & 3.20 $\times$ 10$^{-4}$ \\
IRS5b & 0.26 & 4.73 $\times$ 10$^{-5}$ & -- \\
\enddata
\end{deluxetable}


\begin{figure}[ht!]
  \includegraphics[width=1.0\linewidth]{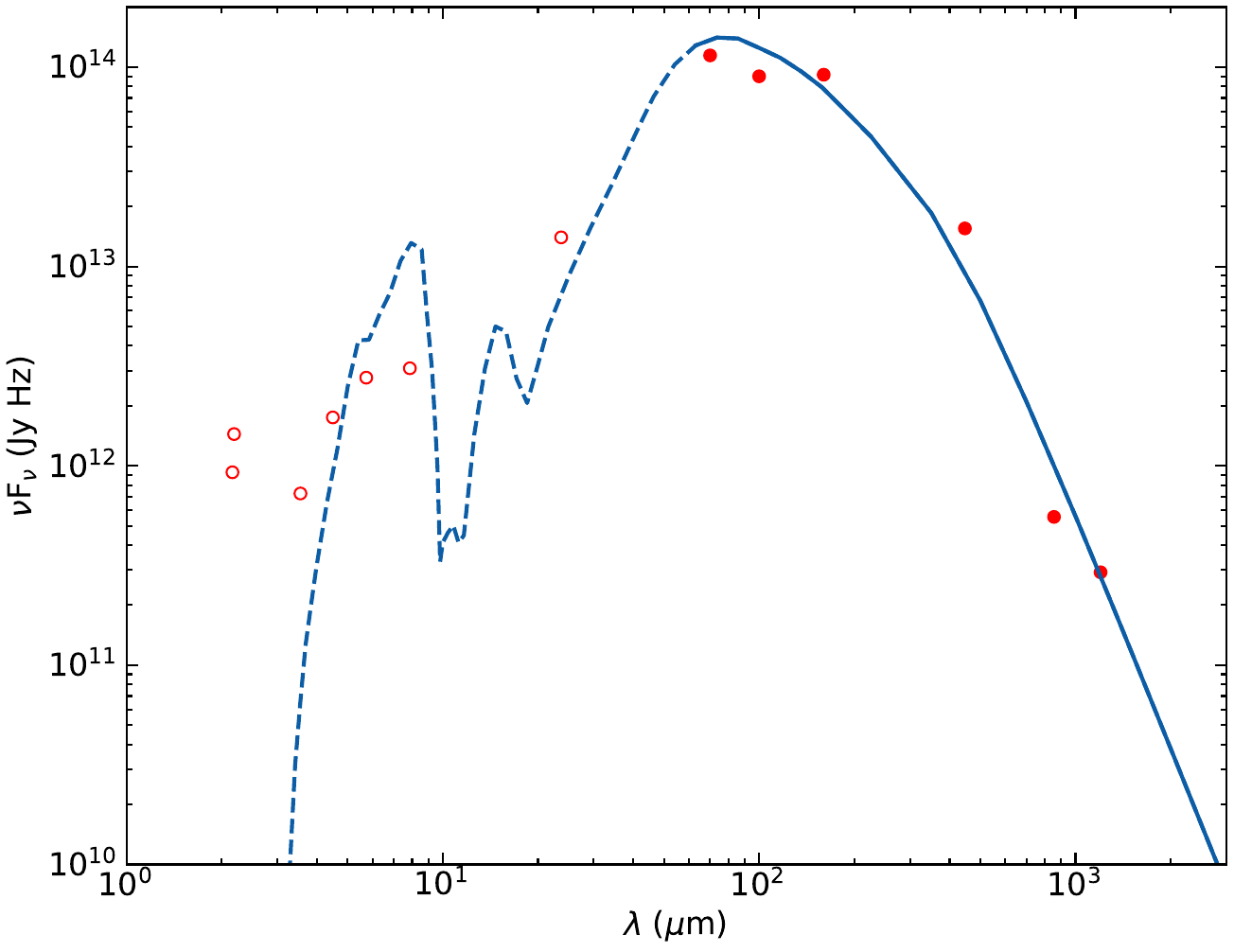} \caption{Fitting of the SED of IRS5N with a 1D radiative transfer model. Filled red circles represent SED values over 60 $\mu$m which are fitted by the model. Open red circles represent the remaining SED values used as input. The solid blue line represents the fit given by the model. The dashed blue line represents that the model does not fit this part of the SED. \label{fig:irs5n_env_mass}}
\end{figure}



For comparison, we estimate the mass of the envelope around IRS5N using a simple 1D dust radiative transfer model. We adopt a single power-law density profile, $n \propto r^{-1.5}$ corresponding to material in free-fall between inner and outer radii of 100 and 10,000~au, respectively, and take the bolometric luminosity of the source determined from the full SED as the sole (internal) heating source of the dust. The dust radiative transfer model then calculates the temperature-profile of the dust in the envelope self-consistently and predicts the SED of the resulting source emission. To constrain the envelope mass we then fit the long wavelength ($\lambda > 60$~$\mu$m) part of the spectral energy distribution of IRS5N. This method allows for a slightly more robust way of determining the envelope mass than simply adopting a single submilimeter flux point and isothermal dust as it provides an estimate of the temperature of the dust taking into account the source luminosity \citep[e.g.,][]{Jorgensen_2002,Kristensen_2012}. The resulting fit of the envelope model is shown in Fig.~\ref{fig:irs5n_env_mass} with the envelope mass constrained to be 1.2~$M_\odot$. The estimated uncertainty on the fitted envelope mass is comparable to the flux calibration uncertainty, typically about 20\% for the measurements used here. However, systematic uncertainties of the adopted simplified physical structure of the envelope and the dust opacity laws will likely dominate over this. It is worth emphasizing that this simplified model is not expected to, and does not, fit the emission at wavelengths shorter than 60~$\mu$m 
due to the complex geometry of the system at small scales and contributions from scattering. 

\subsection{Molecular lines}
\begin{figure*}[ht!]
    \includegraphics[trim={0cm 0cm 1cm 0cm},clip, width=\textwidth, height=\textheight, keepaspectratio]{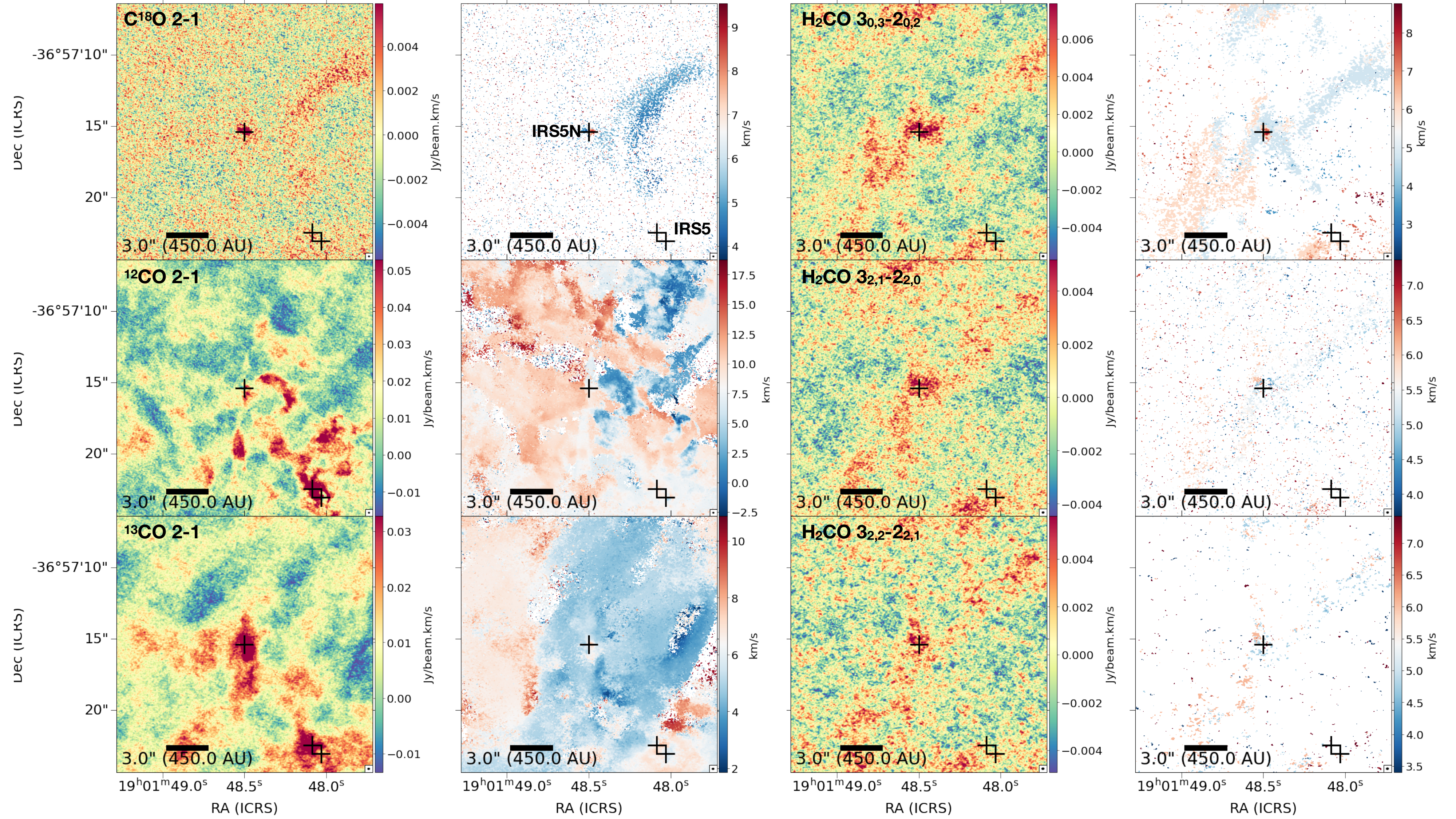}
\caption{Overview of all the molecular lines detected towards IRS5N and IRS5. For each molecule, the moment 0 map is shown on the left and the moment 1 map is shown on the right. The cross marks show the peak of the continuum emission of IRS5N and the binary IRS5. The maps are created by integrating over the velocity ranges of 3.68 -- 9.53 km s$^{-1}$, -5.38 -- 20.65 km s$^{-1}$, 0.51 -- 12.03 km s$^{-1}$, 2.06 -- 8.76 km s$^{-1}$, 3.68 -- 7.36 km s$^{-1}$, and 3.40 -- 7.42 km s$^{-1}$ for \C18O, \tlvco, \thrco, H$_2$CO (3$_{0, 3}$--$2_{0, 2}$), H$_2$CO (3$_{2, 1}$--$2_{2, 0}$), and H$_2$CO (3$_{2, 2}$--$2_{2, 1}$) respectively.} The synthesized beam is shown in black at the bottom right corner of each image, enclosed by a square. \label{fig:lines}
\end{figure*}

Among the molecules mentioned in Section~\ref{sec:data}, emission is detected in \C18O, \tlvco, \thrco, and H$_2$CO molecules in our observations. Figure~\ref{fig:lines} presents an overview of the integrated-intensity (moment-0) and mean-velocity (moment-1) maps of all the detected molecules toward IRS5N and IRS5. The moment 1 maps were generated by integrating the regions where $I_{\nu}\geq3\sigma$, where $\sigma$ is the rms per channel. The maps for \C18O and \tlvco~were made using a robust parameter of 0.5, while the maps for the remaining molecules were made using a robust parameter of 2.0. The channel maps of all the observed molecules around IRS5N are shown in Appendix~\ref{appendix:lines}. 

It is worth emphasizing that large-scale negative components are visible in the channel maps of the molecules, particularly of the CO isotopologues. These negative components indicate that a significant amount of extended flux originating from the large-scale structures surrounding the sources is being resolved out. While it is crucial to analyze these structures to build a comprehensive picture of the physics and chemistry of the system, we are constrained by the limitations of our high-resolution observations. The maximum recoverable scale, ($\theta_{MRS}$) of our observations was 2\farcs91. Hence, this study focuses only on small-scale structures, such as the disk and envelope of individual systems.

\subsubsection{\C18O}
\begin{figure*}[ht!]
    \includegraphics[width=0.52\linewidth]{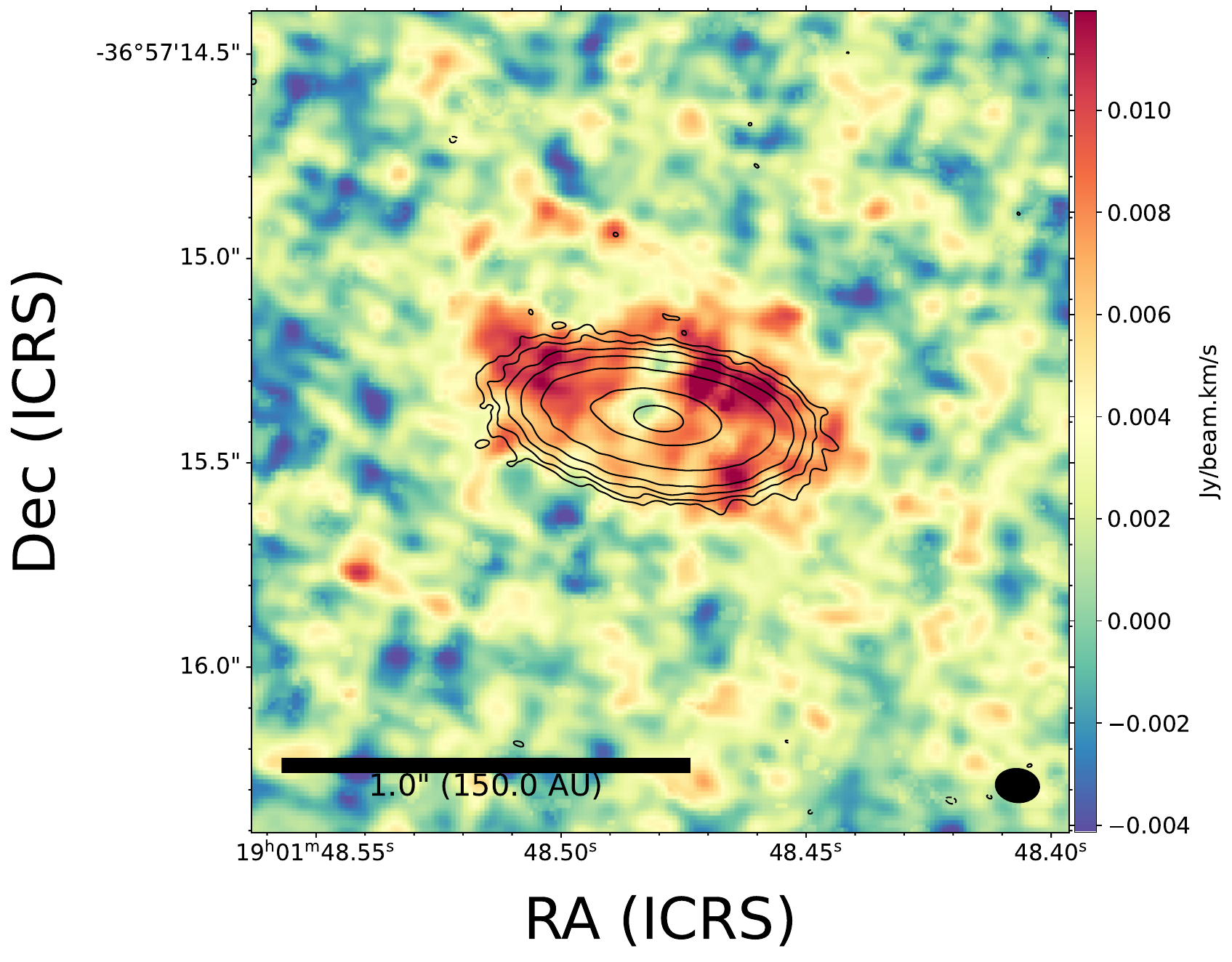}
    \includegraphics[width=0.49\linewidth]{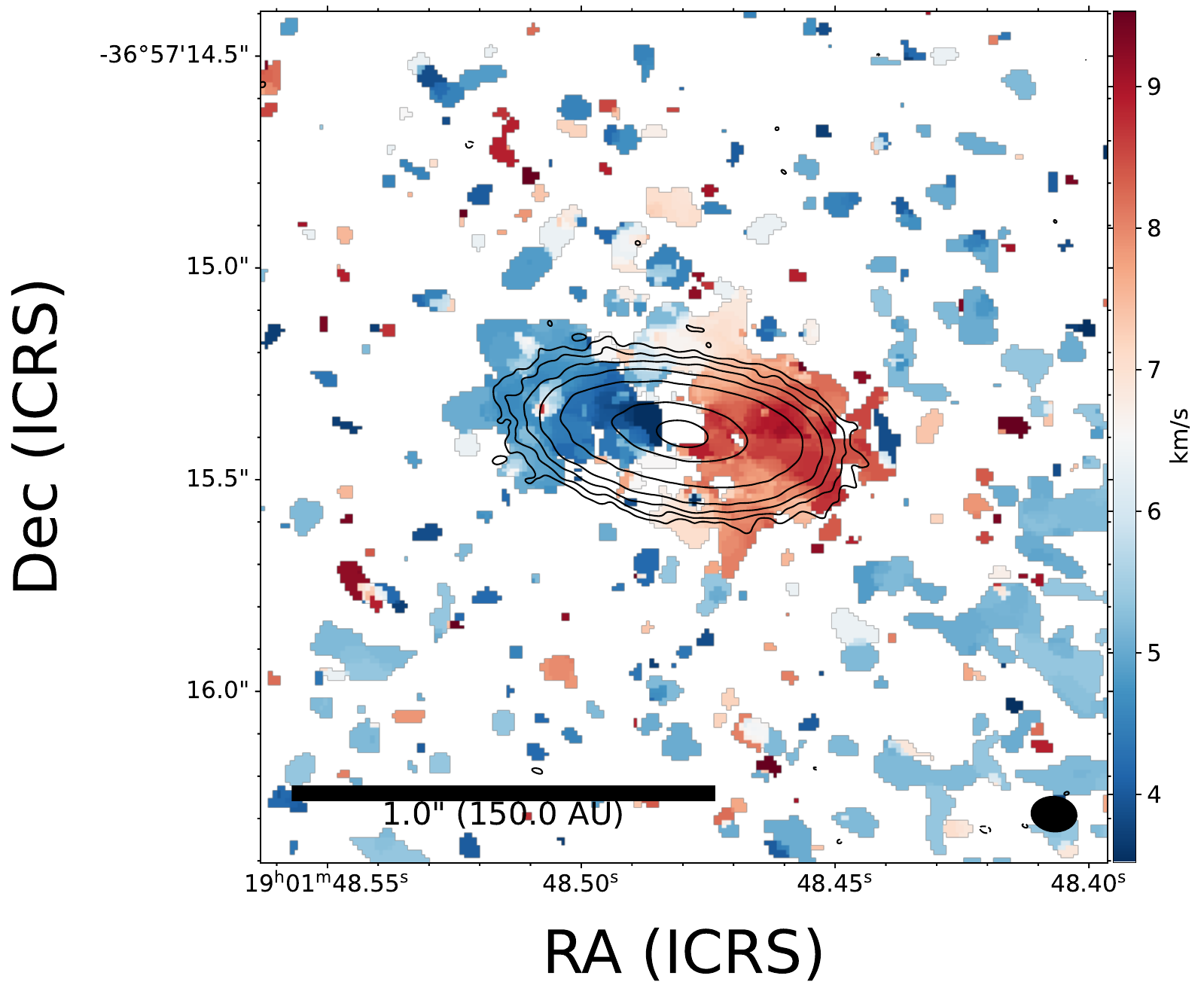}
\caption{Zoomed-in moment 0 map (\textit{left}) and moment 1 map (\textit{right}) of the \C18O ($J$=2--1) emission towards IRS5N created with the robust value of 0.5. The overlaid contours show continuum emission from 3$\sigma$ to 192$\sigma$, with each contour doubling the previous $\sigma$ level. The maps are created by integrating over the velocity ranges of 3.68 -- 9.53 km s$^{-1}$. The synthesized beam size for the \C18O emission is 0$\farcs$11 $\times$ 0$\farcs$08. \label{fig:c18o_irs5n}}
\end{figure*}

Figure~\ref{fig:c18o_irs5n} shows the zoomed-in integrated moment 0 and moment 1 maps of the \C18O (2--1) emission around IRS5N. The moment 0 map shows a flattened structure along with a velocity gradient extended along the major axis of the disk, traced by the continuum emission. The size of the gas disk radius from the \C18O emission is comparable to that of the disk continuum and has a hole at the protostar position. Based on the consistency between the \C18O emission and the continuum emission, the radius of the disk can be assumed to be the same as the FWHM of the continuum, $\sim$62 au.
The hole at the protostellar position has negative intensities below 3$\sigma$ at 5.35 km s$^{-1}$ -- 6.02 km s$^{-1}$ (see Figure~\ref{fig:appendix_c18o_channel}, \ref{fig:appendix_c18o_channel_zoomed}). The deficit likely results from continuum over-subtraction due to the \C18O emission being relatively weak compared to the bright continuum emission.

The moment 1 map of the \C18O emission shows that the blue- and the red-shifted velocities have a distinct separation along the eastern and western sides, respectively. Such a velocity profile is consistent with a rotating disk. The position-velocity (PV) analysis of the \C18O emission is presented in Section~\ref{sec:discussion}.

\subsubsection{\tlvco~and \thrco}

\begin{figure*}[ht!]
    \includegraphics[trim={0cm 0cm 0cm 0cm},clip, width=1\linewidth]{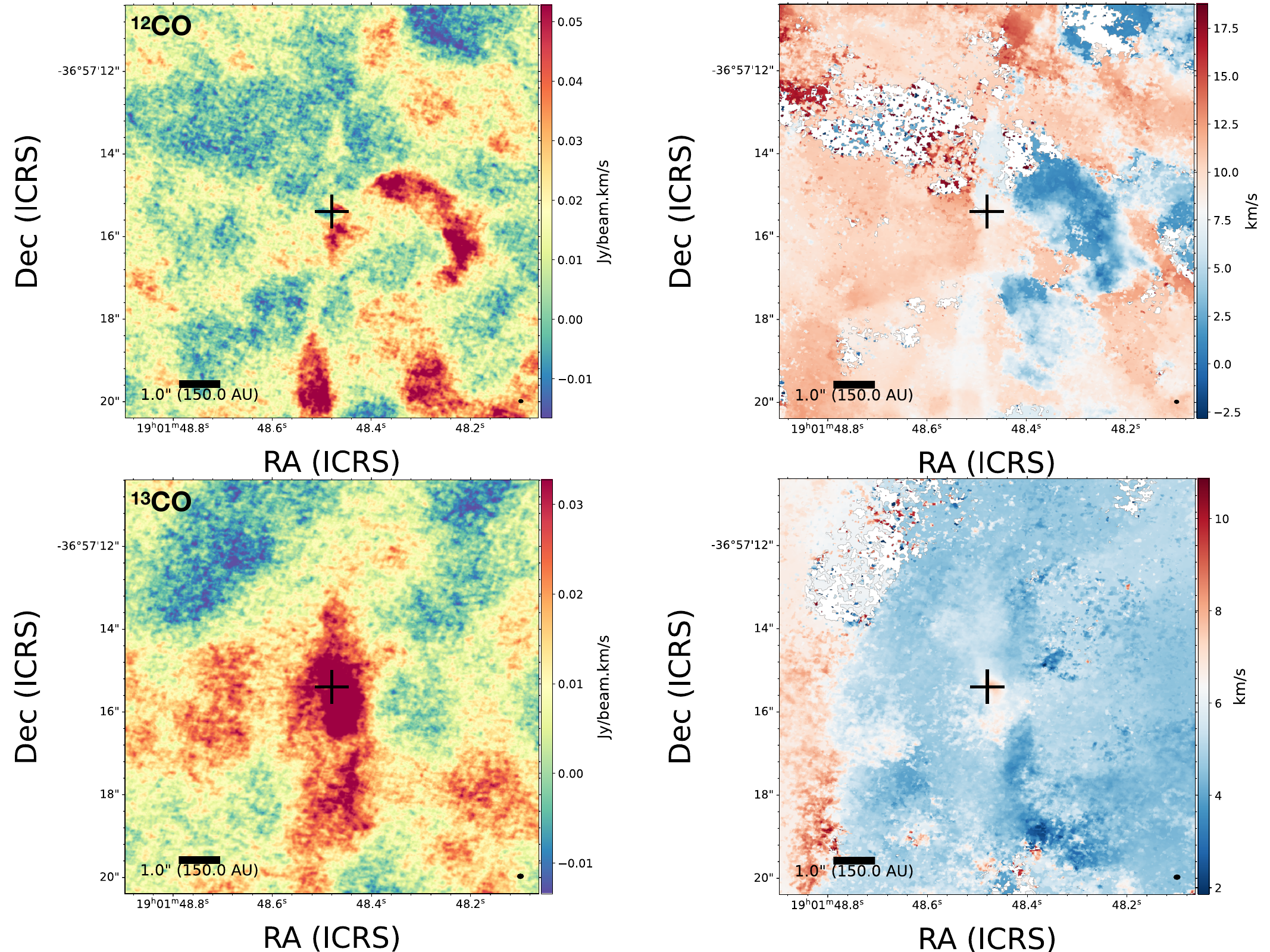}\caption{Zoomed-in moment 0 (\textit{left}) and moment 1 (\textit{right}) maps of \tlvco~(\textit{top}) and \thrco~(\textit{bottom}) emission. The maps are created by integrating over the velocity ranges of -5.38 -- 20.65 km s$^{-1}$ and 0.51 -- 12.03 km s$^{-1}$ for \tlvco~and \thrco~, respectively. The cross represents the peak position of the IRS5N continuum. Synthesized beam is shown in black at the bottom right corner of each image. \label{fig:12co_13co_irs5n}}
\end{figure*}

Figure~\ref{fig:12co_13co_irs5n} shows zoomed-in moment 0 and moment 1 maps of \tlvco~(2--1) and \thrco~(2--1) emission near IRS5N. While the \tlvco~emission shows extended emission around IRS5N, it does not seem to trace any obvious outflow/jet associated with the protostar, which is puzzling. The spiral structure seen towards the west of the protostar is blue-shifted and seems to trace infalling material onto the protostellar disk (see channel maps; Figure~\ref{fig:appendix_12co_channel}). Additionally, extended emission is seen in the surrounding of IRS5N, some of which likely originates from the protostar. In contrast, the \thrco~plot shows some emission in the north-south direction of the protostar but this emission is mostly observed in red-shifted velocity channels (Figure~\ref{fig:appendix_13co_channel_zoomed}). The channel maps of \thrco~also show an apparent deficit near the protostellar position at the velocity range of 5.02 km s$^{-1}$ -- 6.19 km s$^{-1}$ which is much more prominent than the deficit seen on the \C18O channel maps (Figure~\ref{fig:appendix_c18o_channel_zoomed}). This is most likely due to the continuum over-subtraction, similar to that of the \C18O emission. This suggests that the \thrco~emission is extended and somewhat optically thick, leading it to become resolved-out as $\theta_{MRS}$ = 2\farcs91. The moment maps of \tlvco~and \thrco~reveal the complex nature of the emission around IRS5N.

\begin{figure*}[ht!]
    \includegraphics[width=0.50\linewidth]{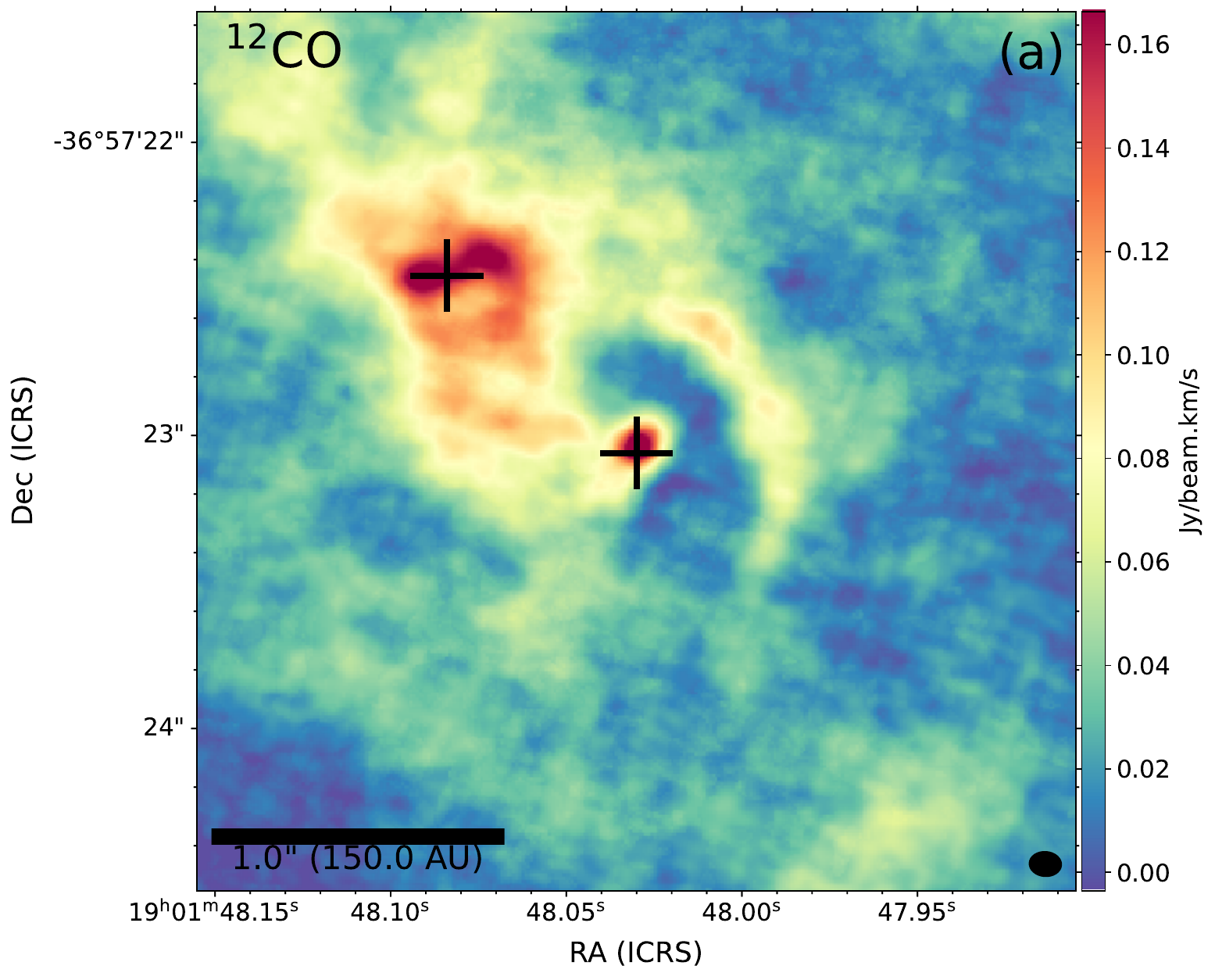}
    \includegraphics[width=0.50\linewidth]{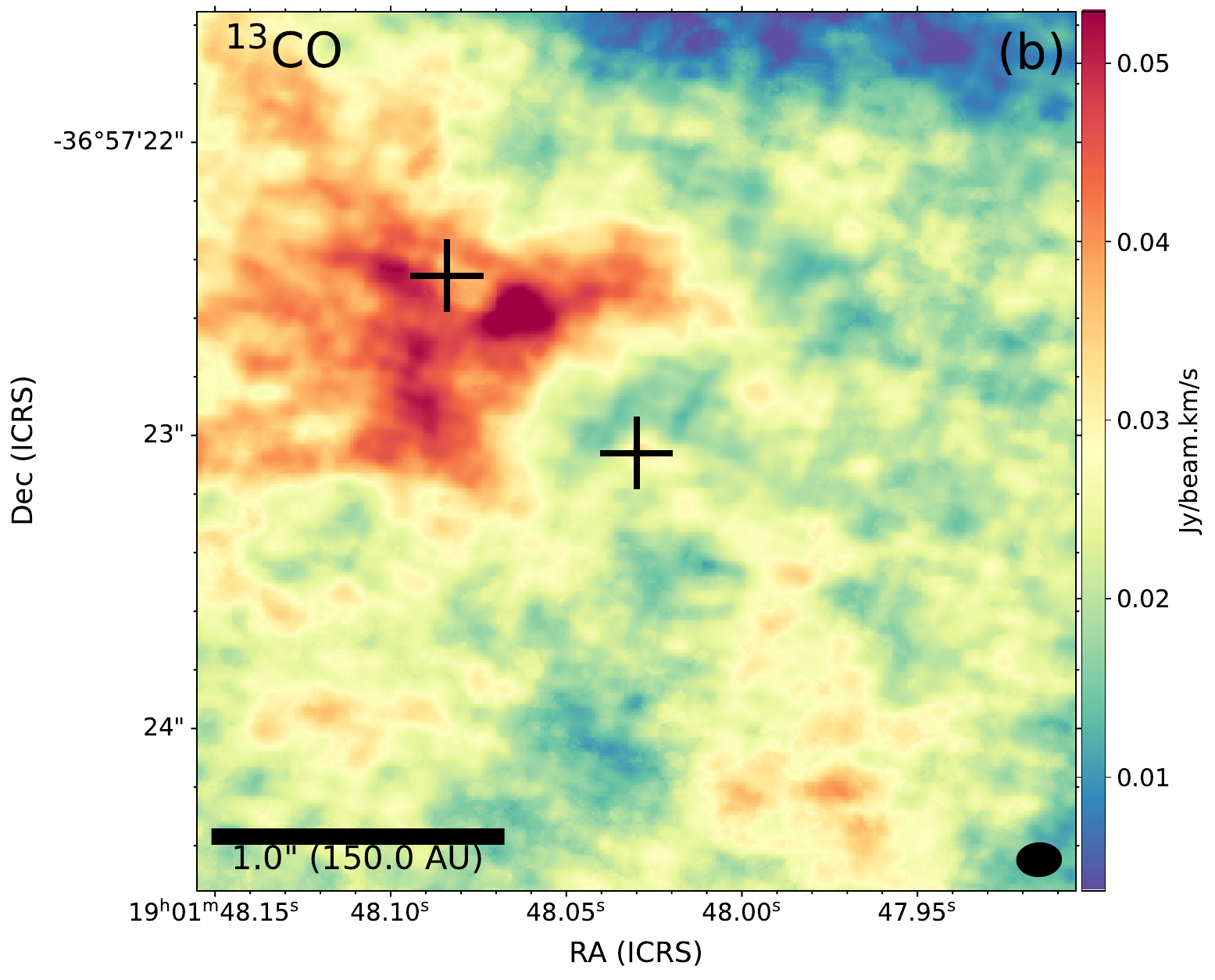}
\caption{Zoomed-in moment 0 maps of \tlvco~emission (a) around the IRS5 binary source and \thrco~emission (b) around the IRS5 binary source. The crosses represent the peak position of the continuum emission of IRS5a and IRS5b sources. The maps are created by integrating over the velocity ranges of -4.75--14.93 km s$^{-1}$ and 2.85 -- 9.86 km s$^{-1}$ for \tlvco~and \thrco~, respectively. Synthesized beam is shown in black at the bottom right corner of each image. \label{fig:12co_13co_IRS5}}
\end{figure*}

\begin{figure*}[ht!]
    \includegraphics[width=1\linewidth]{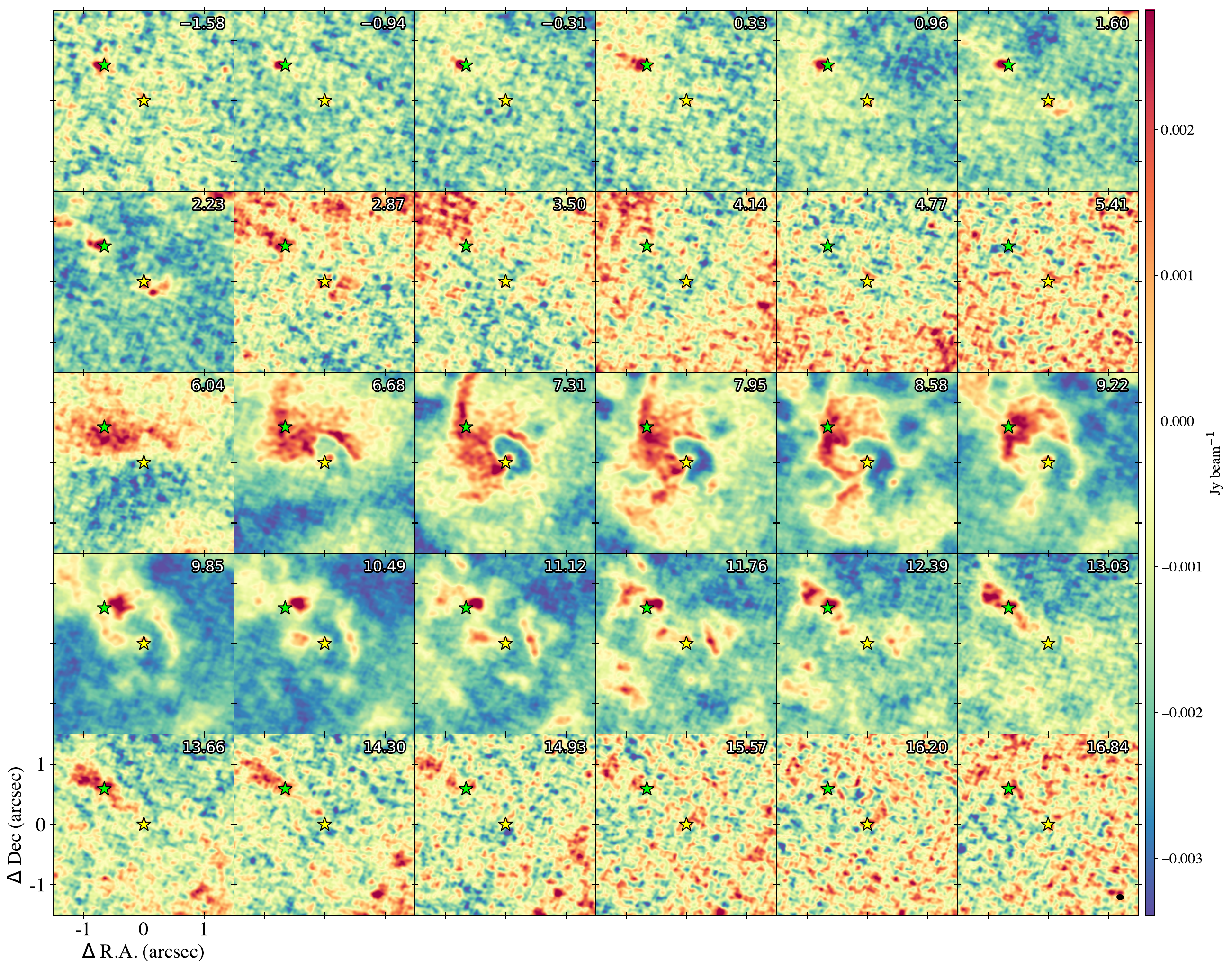}
\caption{Channel maps showing the \tlvco~emission around the two sources of the IRS5 system. The yellow star and the green star show the peak position of the continuum emission of IRS5a and IRS5b, respectively. The numbers at the top show the corresponding velocity of each channel map. Synthesized beam is shown in black at the bottom right corner of the final channel map. \label{fig:12co_channel}}
\end{figure*}

We also detected molecular emission in \tlvco~(2--1) and \thrco ~(2--1) towards IRS5. Figure~\ref{fig:12co_13co_IRS5} (a) shows the moment 0 maps of \tlvco~emission around IRS5a and IRS5b. The \tlvco~emission around IRS5a is compact, with no visible outflow structure. In contrast, bright, elongated emission is observed toward IRS5b in the east-west direction, possibly tracing an outflow from IRS5b. The emission has a velocity gradient and relatively high velocities from -1.58 km s$^{-1}$ -- 2.23 km s$^{-1}$ and 9.85 km s$^{-1}$ -- 13.03 km s$^{-1}$ as shown in the \tlvco~channel maps in Figure~\ref{fig:12co_channel}.
Additionally, bright extended emission is also seen around IRS5b which connects its way into IRS5a. Based on the \tlvco~emission towards the IRS5 binary, we estimate its systemic velocity to be $\sim$6.50 km s$^{-1}$. The channel maps show that the blue-shifted emission seems to emanate from IRS5b and stream onto IRS5a as the velocity increases. Similar stream- or bridge-like features are observed toward other protostellar binaries \citep[e.g.,][]{Sadavoy_2018,van-der-Weil_2019,Jorgensen_2022} and may trace transport of material between the companions triggered by interactions during their mutual orbits \citep[e.g.,][]{Kuffmeier_2019,Jorgensen_2022}. The streaming emission appears to end in a disk-like structure around IRS5a, seen in channel maps of 6.68 km s$^{-1}$ -- 9.85 km s$^{-1}$. Notably, this structure is much larger than the observed size of the dust continuum structure of IRS5a seen in Figure~\ref{fig:cont_irs5n}, indicating it likely traces the inner envelope surrounding the disk. Additionally, in channel maps ranging from 7.31 km s$^{-1}$ to 8.58 km s$^{-1}$, extended emission possibly tracing an outflow is seen towards the southeast of IRS5a. Conversely, the \thrco~emission in Figure~\ref{fig:12co_13co_IRS5} (b) traces the extended mutual envelope material surrounding both sources. The emission is much brighter towards IRS5b than IRS5a, with the brightness peak towards the southwest of IRS5b.

\subsubsection{H$_2$CO}

\begin{figure*}[ht!]
    \includegraphics[width=0.48\linewidth]{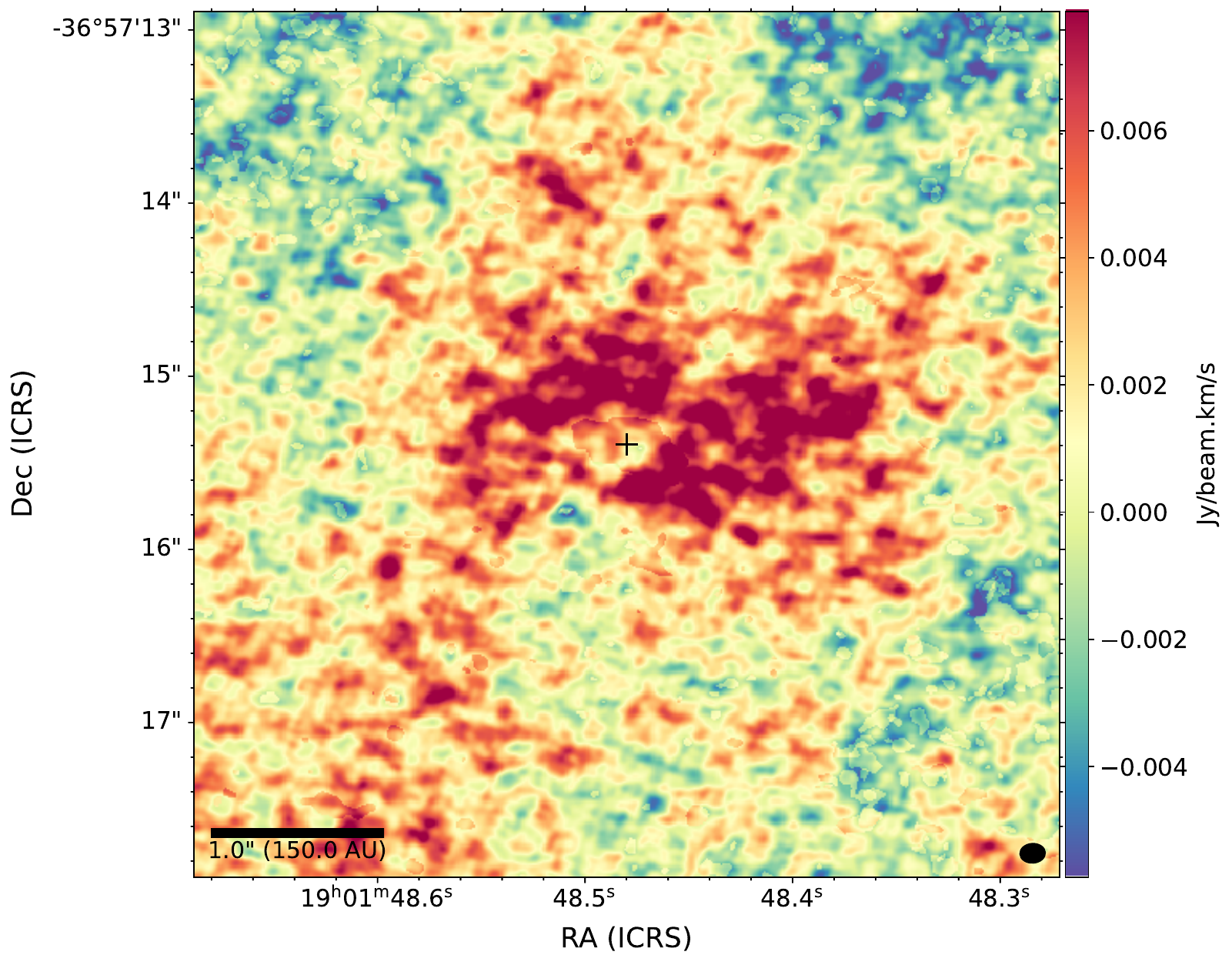}
    \includegraphics[width=0.45\linewidth]{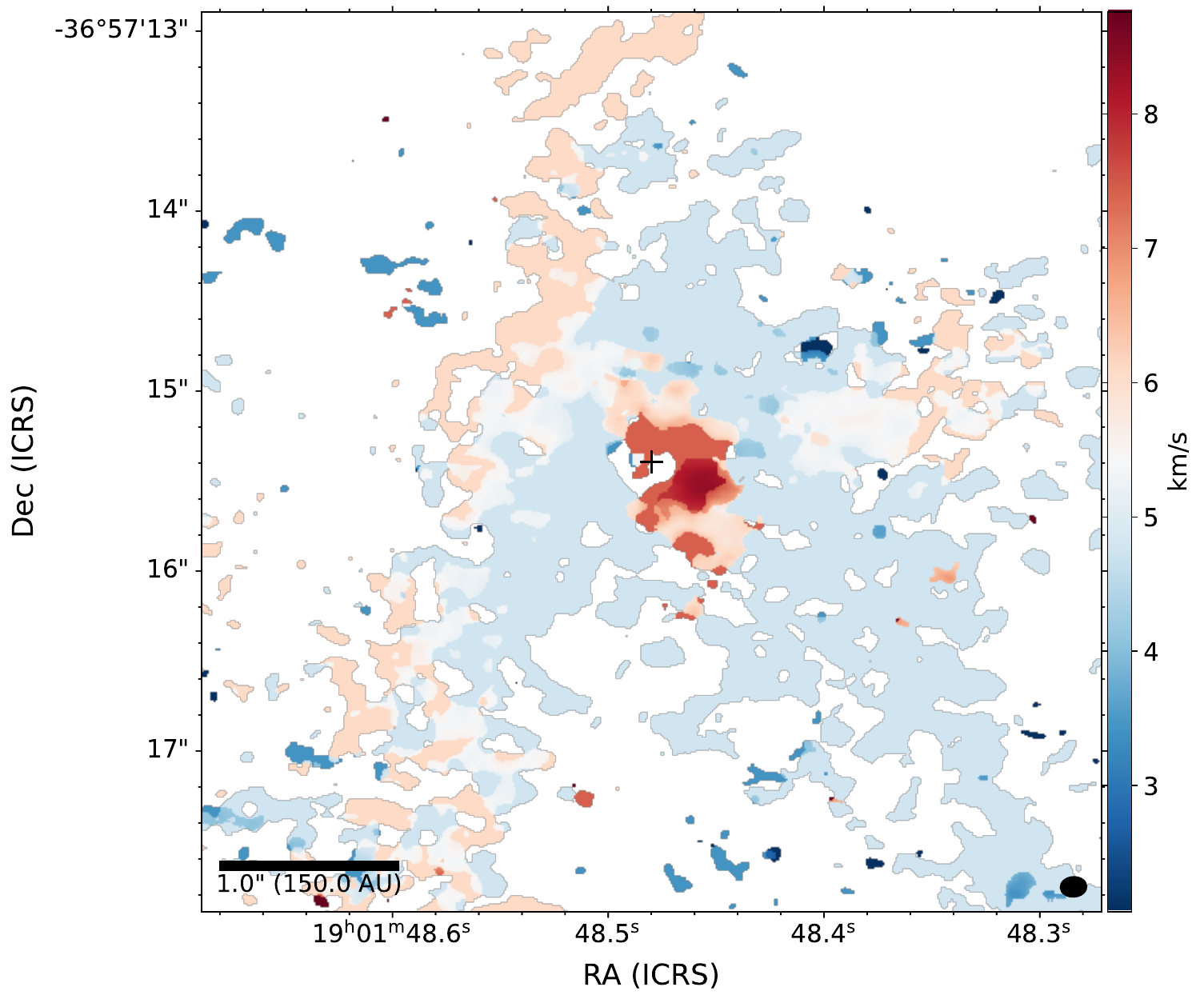}
\caption{Same as Figure~\ref{fig:12co_13co_irs5n} but for H$_2$CO (3$_{0, 3}$--$2_{0, 2}$) instead. The maps are created by integrating over the velocity ranges of 2.06 -- 8.76 km s$^{-1}$.\label{fig:h2co_IRS5n}}
\end{figure*}

In addition to the CO isotopologues, we also detect emission from three H$_2$CO lines towards IRS5N. 
Figure~\ref{fig:lines} shows that the emission structure of the three transitions are similar to one another, with most of the emission surrounding the disk and inner envelope with extended emission towards the northwest and southeast direction of the source. There is a slight velocity difference between the two sides of the extended source as shown by the moment 1 maps. The 3$_{0, 3}$--$2_{0, 2}$ transition has the lowest upper-level energy and is also the strongest, as expected. A magnified view of the moment 0 and moment 1 maps of the brightest transition of H$_2$CO, $3_{0, 3}$--2$_{0, 2}$, is shown in Figure~\ref{fig:h2co_IRS5n}. The zoomed-in maps reveal that besides the large-scale emission, some red-shifted emission is visible towards the west of the disk, similar to that of the \C18O~emission (see Figure~\ref{fig:c18o_irs5n}), but appears to lack the corresponding blue-shifted counterpart, suggesting asymmetric distribution of the chemical composition of the disk/envelope system.
The velocity channel maps show that there is negative emission at the position of the protostar which again is likely caused by continuum over-subtraction (see Figure~\ref{fig:appendix_h2co_3-03_2-02_channel}, \ref{fig:appendix_h2co_3-03_2-02_channel_zoomed}). However, the large-scale negatives seen in the velocity channel maps suggest that a significant amount of flux is getting resolved out.


\section{Analysis and Discussion}\label{sec:discussion}
\subsection{Continuum Modeling}

As shown in Figure~\ref{fig:cont_irs5n}, even though we sufficiently resolve the disk of IRS5N, no apparent substructures can be identified in the continuum emission. IRS5a also appear to be relatively smooth, while IRS5b is not resolved.

\begin{figure*}[ht!]
\includegraphics[width=0.48\linewidth]{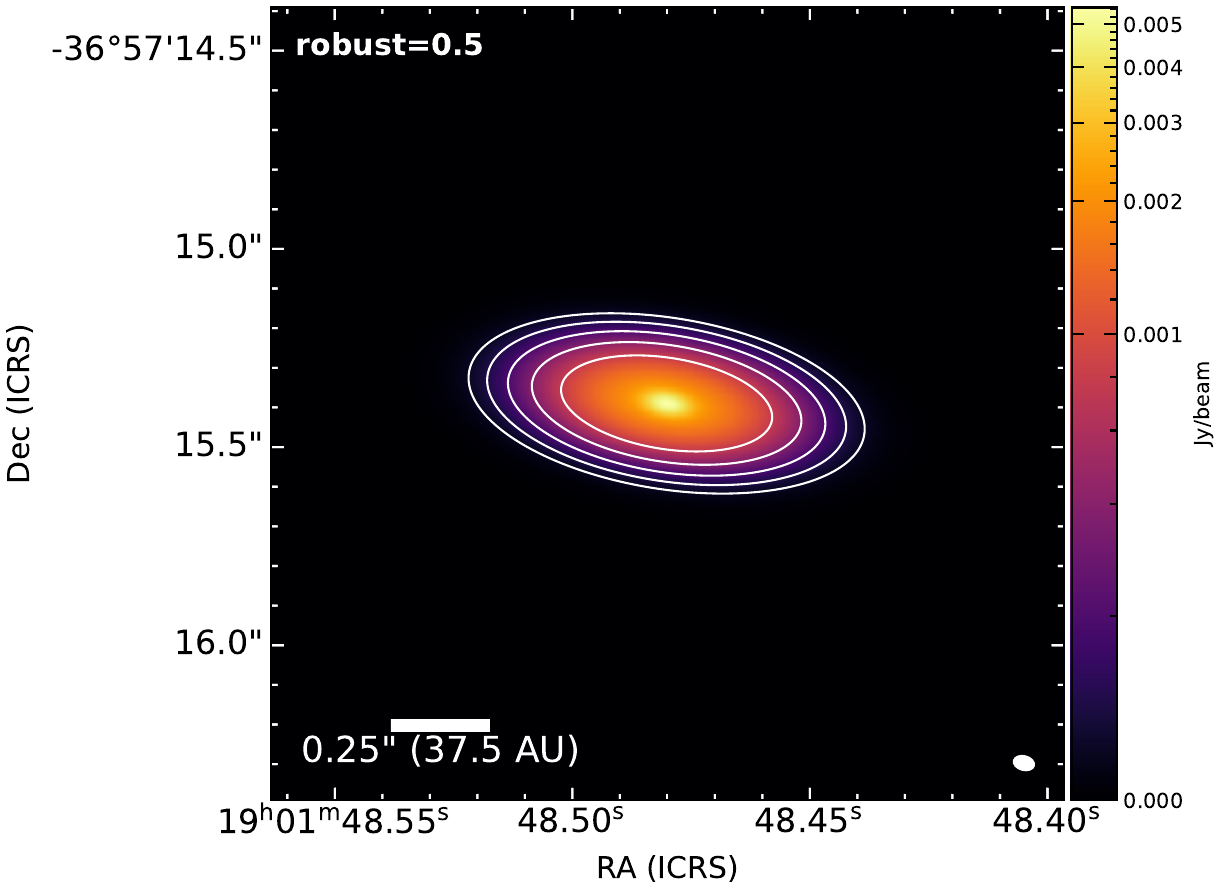} 
\includegraphics[width=0.5\linewidth]{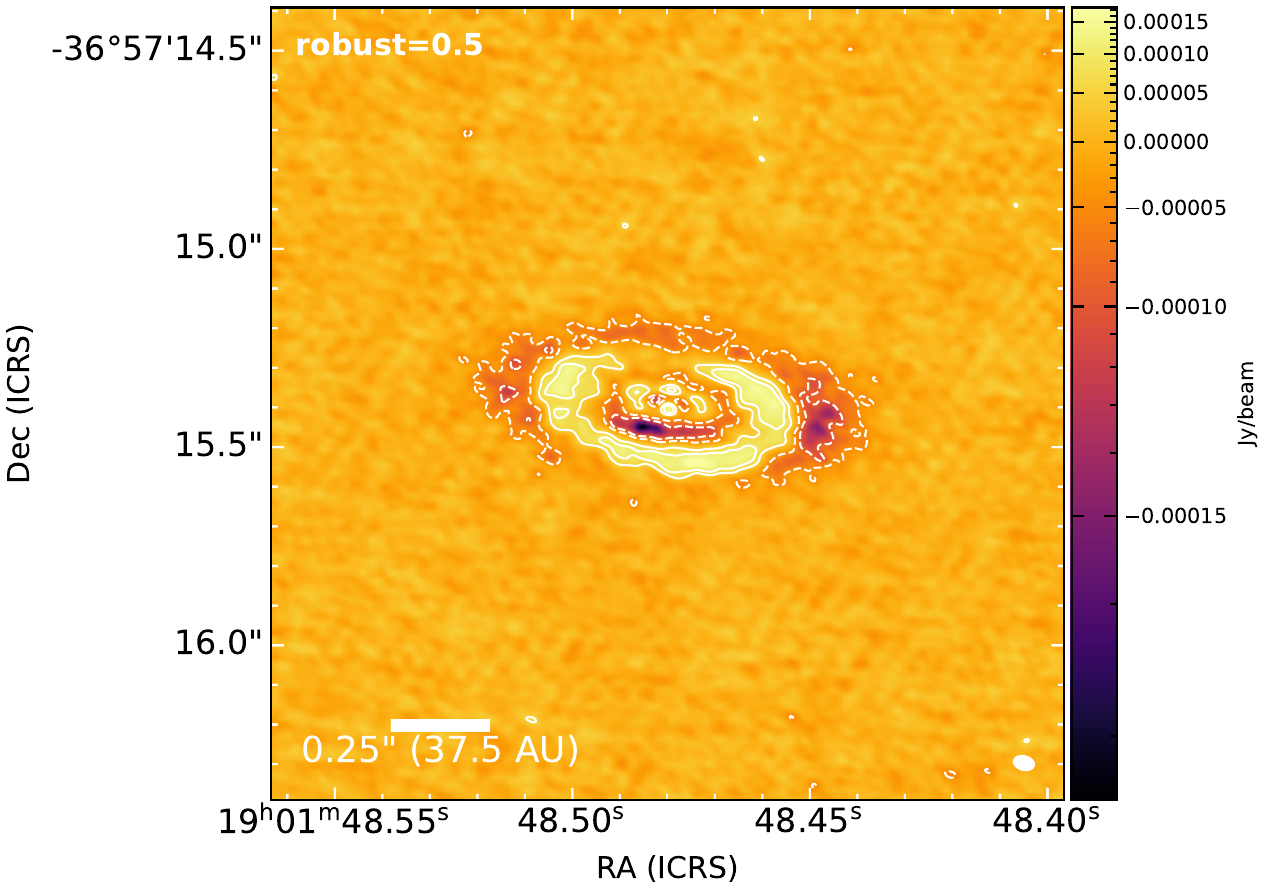}
  \caption{IRS5N disk model (\textit{left}) and residual (\textit{right}) created using double Gaussian components. Contours show the 3$\sigma$, 6$\sigma$, 12$\sigma$, 24$\sigma$ and 48$\sigma$. The dashed contours indicate negative intensity. The white ellipse on the bottom right corner of each image shows the synthesized beam. \label{fig:model_irs5n_2peaks}}
\end{figure*}

\begin{figure}[ht!]
  \includegraphics[width=1.0\linewidth]{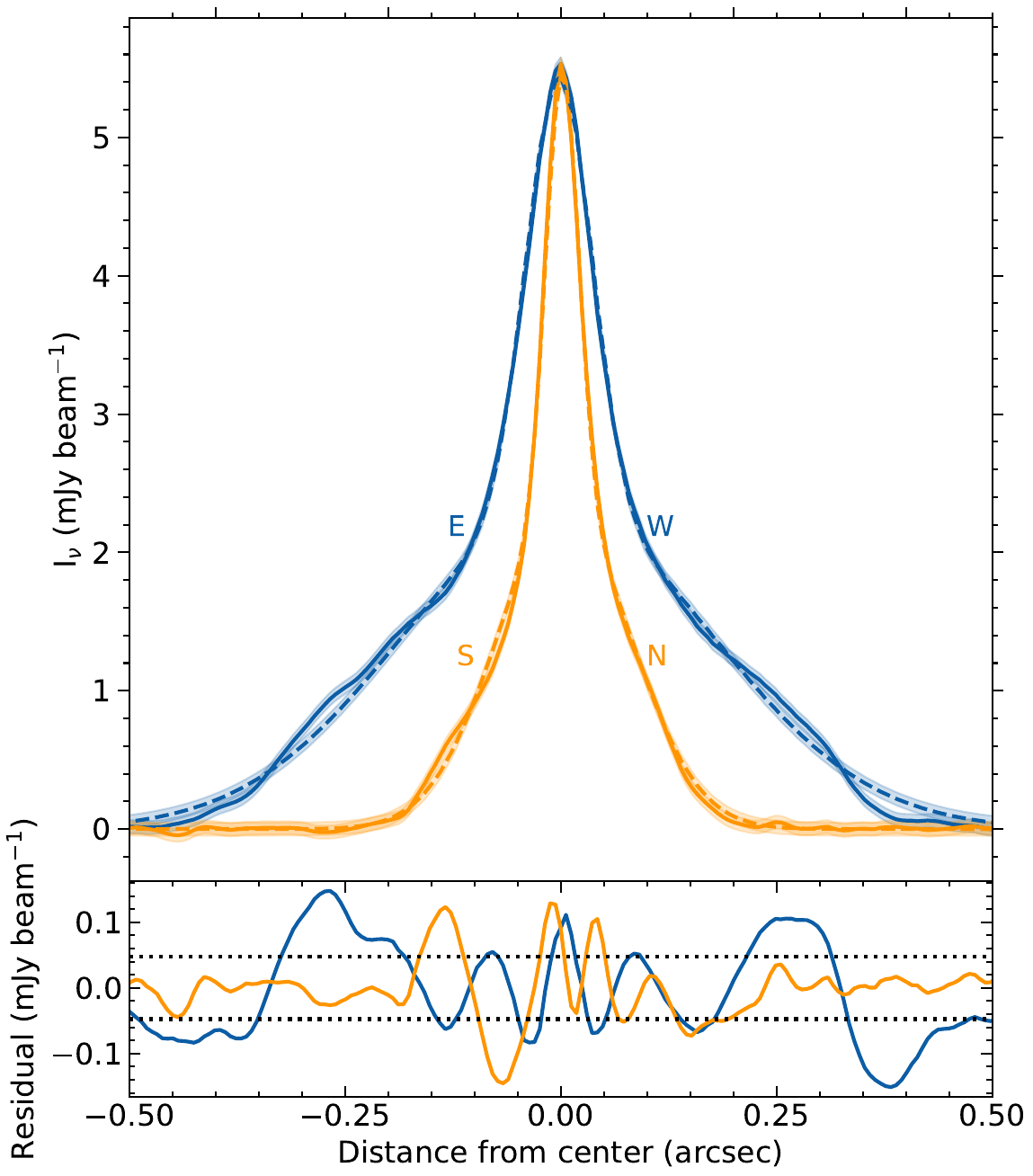} \caption{Intensity profiles of the 1.3 mm continuum emission of IRS5N shown in Figure~\ref{fig:cont_irs5n}(b) and double Gaussian component model shown in Figure~\ref{fig:model_irs5n_2peaks}. The solid line represents the observed emission and the dashed line represents the model. The blue lines indicate the intensity along the major axes and the orange lines represent the intensity along the minor axes. The smaller plot at the bottom shows the residual intensity after subtracting the model from the observation. The shaded region and the horizontal dotted lines indicate $\pm3\sigma$ uncertainties with $\sigma = 0.016$ mJy beam$^{-1}$. \label{fig:irs5n_intensity2}}
\end{figure}

Figure~\ref{fig:model_irs5n_2peaks} shows the best-fit model and its corresponding residual of the continuum emission of IRS5N made with CASA task \texttt{imfit}. The model was created using two 2D Gaussian components as a single Gaussian model misses a lot of emission of the continuum. 
The residual image and the intensity plots in Figure~\ref{fig:irs5n_intensity2} show that the double-component model is able to recover most of the continuum emission. The fitting results of both models are provided in Table~\ref{tab:fitting}. The parameters of the disk continuum such as its peak position, P.A., and $i$ do not change significantly between the two components of the model. It is important to note that the residuals are a result of the model not representing the structure of the emission and can not necessarily be taken as evidence of the presence of substructures in the distribution of material within the dusty disk.  The residual image shows that there is some asymmetry in the direction of the minor axis (North-South). The disk appears to be brighter in the south compared to the north. Such asymmetry in the minor axis is observed in several eDisk sources \citep{Ohashi_2023}. This can be attributed to the geometrical effects of optically thick emission and flaring of the disk \citep{Takakuwa_2023}. The north side of the disk is more obscured compared to the south which is expected to be on the far side of the disk with $i\sim 65^{\circ}$, where 90$^{\circ}$ represents the completely edge-on case.

\begin{figure*}[ht!]
  \includegraphics[width=0.5\linewidth]{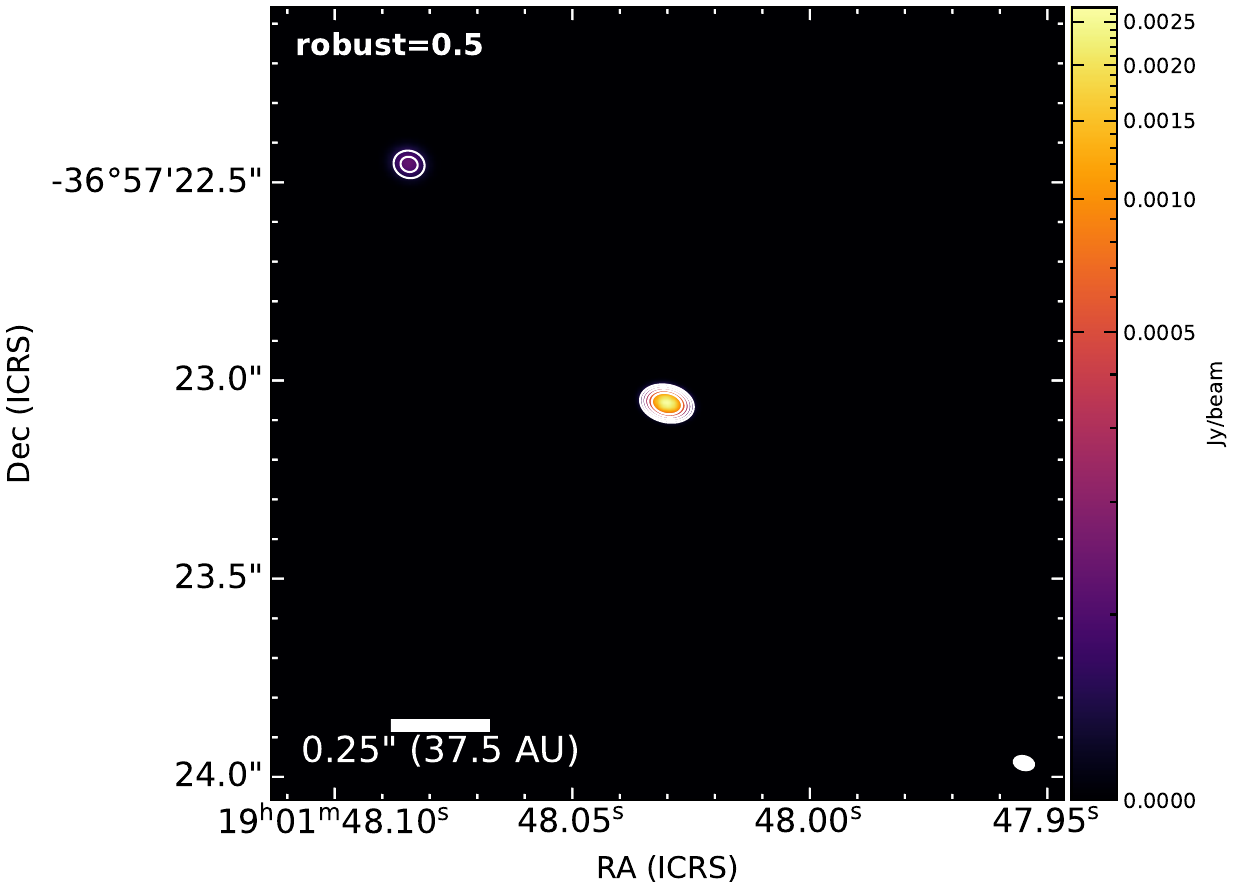}
  \includegraphics[width=0.5\linewidth]{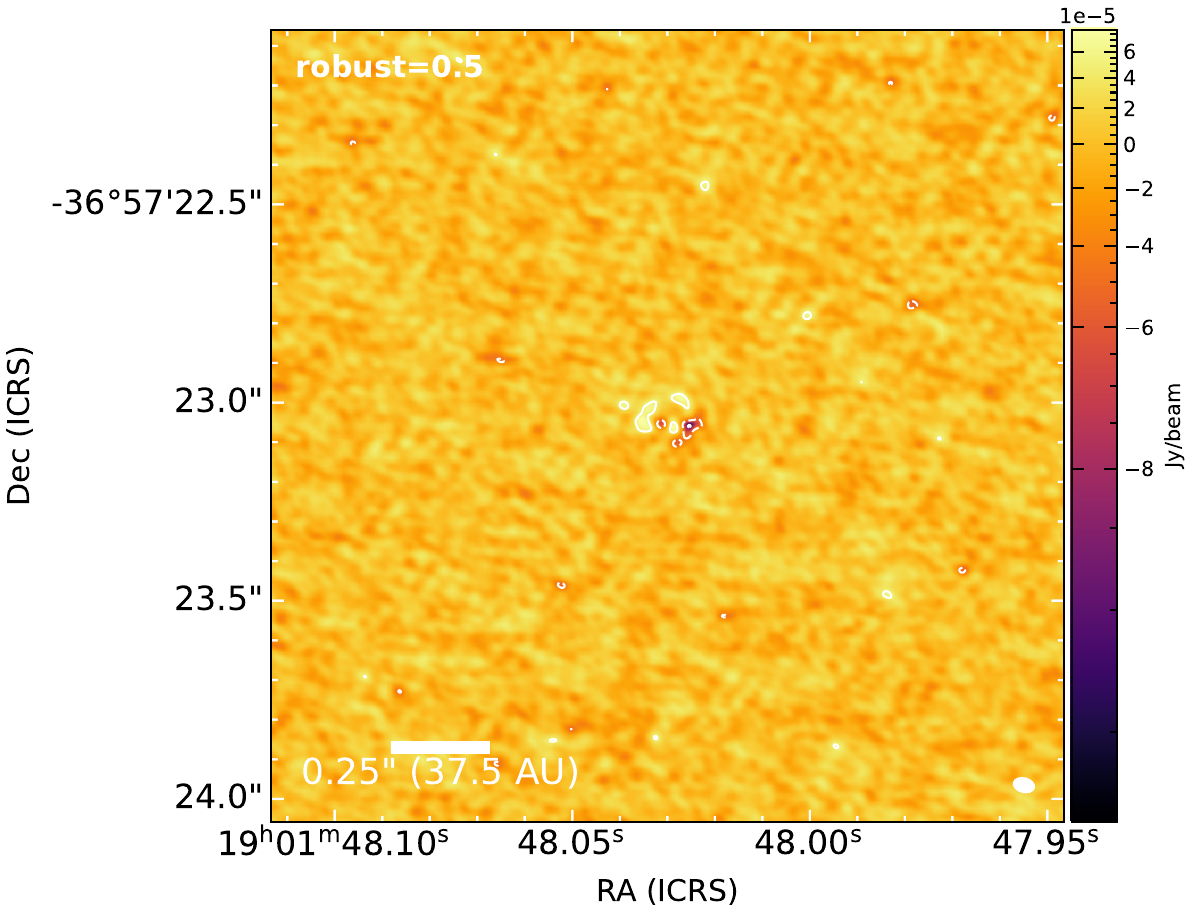}
  \caption{Same as Figure~\ref{fig:irs5n_intensity2} but for IRS5a binary source with the model made with a single Gaussian peak for each continuum emission observed. \label{fig:model_irs5}}
\end{figure*}

We also fit the continuum emission for both sources in the IRS5 system. Figure~\ref{fig:model_irs5} shows the model and the residual created of IRS5a and IRS5b after subtracting a single 2D Gaussian. The model is able to reasonably capture the majority of the continuum emission from both sources as seen from the residual images. The results of the fitting are provided in Table~\ref{tab:fitting}. 

\begin{deluxetable*}{cccccc}
\tablecaption{Best-fit parameters. \label{tab:fitting}}
\tablehead{
\colhead{Parameter} & \multicolumn{3}{c}{Single Gaussian Component} & \multicolumn{2}{|c}{Double Gaussian Components} \\
\colhead{} & \colhead{IRS5N} & \colhead{IRS5a} & \colhead{IRS5b} & \multicolumn{2}{|c}{IRS5N} \\
\colhead{} & \colhead{} & \colhead{} & \colhead{} & \multicolumn{1}{|c}{Component 1} & \colhead{Component 2} 
}
\startdata
R.A (h:m:s) & 19:01:48.480 & 19:01:48.030  & 19:01:48.084 & \multicolumn{1}{|c}{19:01:48.480} & 19:01:48.480 \\
Dec (d:m:s) & -36:57:15.39 & -36:57:23.06 & -36:57:22.46 & \multicolumn{1}{|c}{-36:57:15.39} & -36:57:15.39 \\
Beam size ($\farc$) & 0.05 $\times$ 0.03 & 0.05 $\times$ 0.03 & 0.05 $\times$ 0.03 & \multicolumn{1}{|c}{0.05 $\times$ 0.03} & 0.05 $\times$ 0.03 \\
Beam P.A. ($^{\circ}$) & 75.44 & 75.44 & 75.44 & \multicolumn{1}{|c}{75.44} & 75.44 \\
$\theta_{\mathrm{maj}}$ (mas) $^{\dagger}$ & 373.7 $\pm$ 7.1 & 22.91 $\pm$ 0.95 & 46 $\pm$ 16 & \multicolumn{1}{|c}{73.58 $\pm$ 1.69} & 423.9 $\pm$ 3.0 \\
$\theta_{\mathrm{min}}$ (mas) $^{\dagger}$ & 156.7 $\pm$ 2.9 & 16.82 $\pm$ 0.56 & 40 $\pm$ 26 & \multicolumn{1}{|c}{28.32 $\pm$ 0.76} & 179.4 $\pm$ 1.3 \\
P.A. ($^{\circ}$) $^{\dagger}$ & 81.10 $\pm$ 0.76 & 84.7 $\pm$ 5.7 & 174 $\pm$ 54 & \multicolumn{1}{|c}{81.22 $\pm$ 0.80} & 81.08 $\pm$ 0.29 \\
Inclination ($^{\circ}$) & 65.21 $\pm$ 0.70 & 42.76 $\pm$ 3.29 & 29.59 $\pm$ 74.38 & \multicolumn{1}{|c}{67.36 $\pm$ 0.84} & 64.96 $\pm$ 0.27 \\
Peak Intensity (mJy beam$^{-1}$) & 2.970 $\pm$ 0.054 & 3.893 $\pm$ 0.020 & 0.174 $\pm$ 0.00022 & \multicolumn{1}{|c}{3.172 $\pm$ 0.039} & 2.286 $\pm$ 0.016 \\
Flux Density (mJy) & 99.1 $\pm$ 1.9 & 4.727 $\pm$ 0.041 & 0.361 $\pm$ 0.00063 & \multicolumn{1}{|c}{7.09 $\pm$ 0.12} & 98.30 $\pm$ 0.69 \\
\enddata
\tablecomments{$\dagger$ Values are deconvolved from the beam.}
\end{deluxetable*}


\subsection{Kinematics of the disk: Position-velocity diagram}
\begin{figure}[ht!]
  \includegraphics[width=1.0\linewidth]{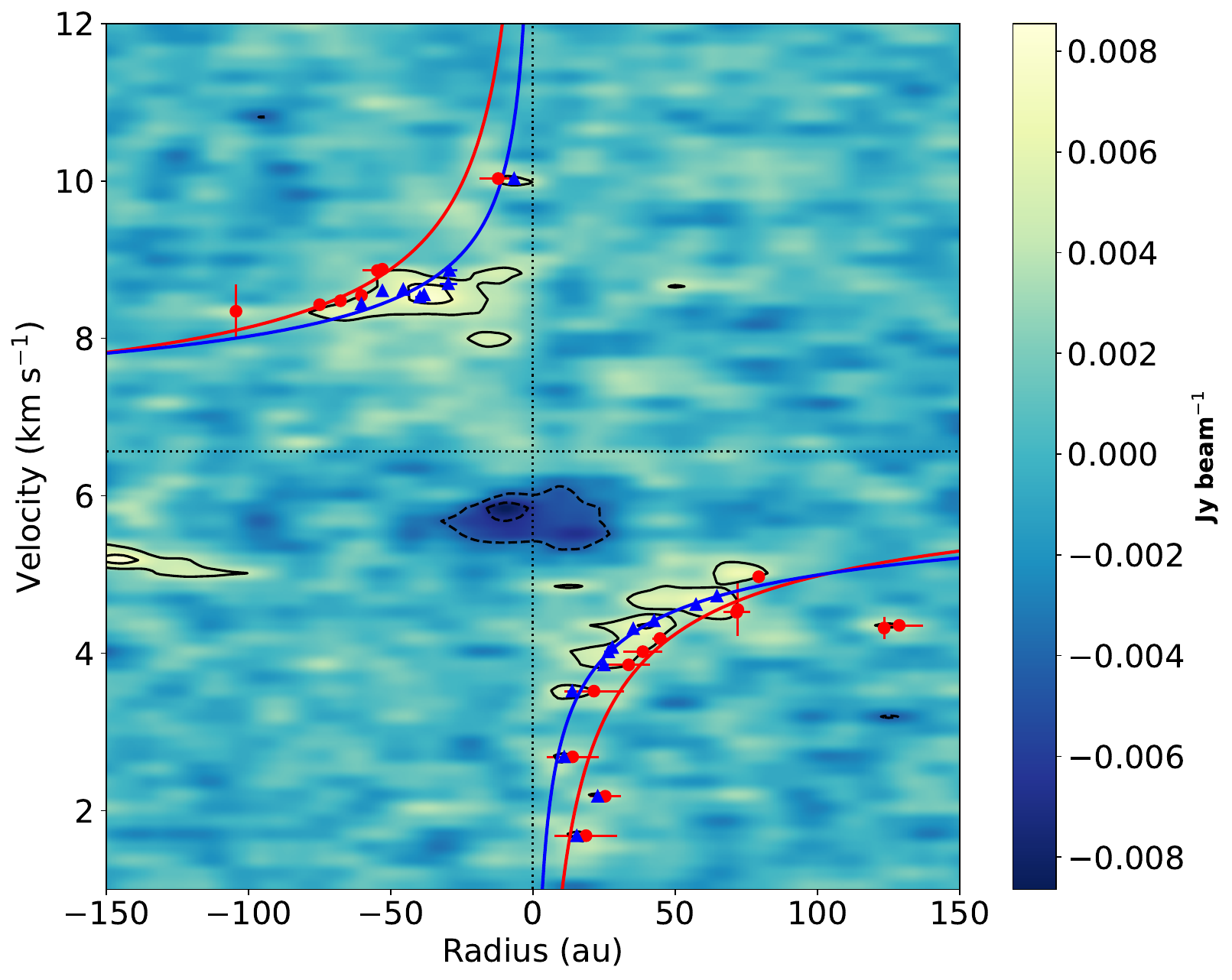} \caption{PV diagram of \C18O emission towards IRS5N cut along its major axis. The points derived using the ``edge" and ``ridge" methods and their corresponding rotation curves are shown in blue and red, respectively. The black contours start at 3$\sigma$ level and increase every 2$\sigma$. Dashed contours show the corresponding negative contours. \label{fig:irs5n_pvdiagram}}
\end{figure}

The kinematics of the protostellar disk are investigated with position-velocity (PV) diagrams of molecular line emission that trace the disk. For IRS5N, \C18O is the only molecule where evidence of rotation is seen in the protostellar disk (see Figure~\ref{sec:results}). \C18O is much less optically thick and is a better disk tracer than other CO isotopologues, making it an excellent species for PV analysis. Figure~\ref{fig:irs5n_pvdiagram} shows the PV diagram of IRS5N in \C18O along the major axis of the disk. The PV diagram shows that the blue-shifted emission and the red-shifted emission are separated in the northeast and the southwest, respectively.

The PV diagram was fitted using the {\tt pvanalysis} package of the Spectral Line Analysis/Modeling (SLAM)\footnote{\url{https://github.com/jinshisai/SLAM}} code \citep{Aso_2023} to investigate the nature of the rotation. The details of the fitting procedure are given in \citet{Ohashi_2023}, but a short description is provided here. The code determines the corresponding position at a given velocity using the PV diagram and calculates two types of representative points known as the edge and the ridge. The ridge is defined as the intensity-weighted mean calculated with emission detected above a given threshold, while the edge corresponds to the outermost contour defined by a given threshold. For the analysis of the PV diagram of the \C18O emission around IRS5N, a threshold of $3\sigma$ level was used, where $\sigma = 1.636$ mJy beam$^{-1}$. The edge and the ridge are then fit separately with a single power-law function given by

\begin{equation}
    V_{\mathrm{rot}} = V_b \left( \frac{R}{R_b} \right)^{-p} + V_{sys},
\end{equation}
where $V_{\mathrm{rot}}$ is the rotational velocity, $R_b$ is the break radius, $V_b$ is the rotational velocity at $R_b$, $p$ is the power-law index, and $V_{sys}$ is the systemic velocity of the system. 

\begin{deluxetable}{ccc}
\tablecaption{PV fitting results for \C18O with SLAM \label{tab:pv_analysis}}
\tablehead{
\colhead{Fitting method} & \colhead{Edge} & \colhead{Ridge} 
}

\startdata
R$_{\mathrm{b}}$ (au) & 76.98 $\pm$ 2.30 & 42.80 $\pm$ 0.44 \\
$p_{\mathrm{in}}$ & 0.554 $\pm$ 0.047 & 0.379 $\pm$ 0.012 \\
$v_{\mathrm{sys}}$ (km s$^{-1}$) & 6.564 $\pm$ 0.024 & 6.507 $\pm$ 0.008 \\
M$_{\mathrm{in}}$ ($M_{\odot}$) & 0.398 $\pm$ 0.041 & 0.184 $\pm$ 0.008 \\
\enddata
\end{deluxetable}

The fitting results of the SLAM code are summarized in Table~\ref{tab:pv_analysis}. Here, the ridge points are calculated using the 1D intensity weighted mean profile, called ``mean" fitting method. However, the ridge points can also be calculated using the center of the Gaussian fitting. Using this ``Gaussian" fitting method, we get R$_{\mathrm{b}}$ = 39.75 $\pm$ 0.76 au, $p_{\mathrm{in}}$  = 0.515 $\pm$ 0.029, $v_{\mathrm{sys}}$ = 6.464 $\pm$ 0.020 km s$^{-1}$, and M$_{\mathrm{in}}$  = 0.246 $\pm$ 0.015 ($M_{\odot}$), which are consistent to the values derived from the ``mean" method. In the case of both the edge and ridge methods, the value of $p_{\mathrm{in}}$ is found to be close to 0.5, suggesting that the disk of IRS5N is already in Keplerian rotation. Typically, Keplerian rotation is commonly observed in more evolved sources \citep{Simon_2000}. However, recent studies have found that some Class 0 sources already possess Keplerian disks \citep[e.g.][]{Tobin_2012,Ohashi_2014,Ohashi_2023}. In both the ridge and the edge methods, Keplerian rotation is observed out to a radius of $\sim$40 au and $\sim$76 au, respectively. The FWHM of the disk continuum falls well within this range, indicating that it could serve as a reliable indicator of the disk size of IRS5N. Under this assumption, the mass of the central source of IRS5N is estimated to be 0.398 $\pm$ 0.041 $M_{\odot}$ and 0.184 $\pm$ 0.008 $M_{\odot}$ for the edge and the ridge cases, respectively. The actual mass of the central source likely lies between these two estimates, approximately 0.3 $M_{\odot}$ \citep{Maret_2020}. This shows that with a stellar mass of $\sim$0.3 $M_{\odot}$ compared to a disk mass of $\sim 0.007 - 0.02 M_{\odot}$ and an envelope mass of 1.2 $M_{\odot}$, IRS5N is a deeply embedded protostar.

The stability of the disk against gravitational collapse can be estimated by using Toomre's $Q$ parameter
\begin{equation}\label{eq:toomre_q}
\centering
Q = \frac{c_s\Omega}{\pi G \Sigma},
\end{equation}
where $c_s$ is the sound speed, $\Omega = GM_*/R^3$ is the differential rotation value of a Keplerian disk at the given radius $R$, $M_*$ is the mass of the protostar, $G$ is the gravitational constant, and $\Sigma$ is the surface density. A disk is considered gravitationally stable if $Q > 1$, while $Q < 1$ suggests that the disk may be prone to fragmentation. This equation can also be expressed in the form given by \citet{Kratter_2016} and \citet{Tobin_2016} as
\begin{equation}\label{eq:toomre_q_approx}
\centering
Q \approx 2\frac{M_* H}{M_d R},
\end{equation}
where $H = c_s/\Omega$, $M_d$ is the mass of the disk, and $R$ is the radius of the disk. For IRS5N with $M_*$ = 0.3 $M_\odot$ and $R = 62$ au, we find $Q \approx 3.5$ and 15 for disk masses of 0.019 $M_\odot$ and $6.65 \times 10^{-3} M_\odot$ at 20 K and 47 K, respectively. This implies that the disk of IRS5N is gravitationally stable.


\subsection{The low molecular emission around IRS5N}
In Section~\ref{sec:results}, we mention that although we see extended emission in \tlvco~and \thrco~in the region around IRS5N, we do not see any clear signs of outflow in these molecules. Emission is also not detected in SiO ($J$=5--4) or SO ($J$=6$_5$--5$_4$), both of which are known tracers of outflow and shocks \citep[e.g.,][]{Schilke_1997,Wakelam_2005,Ohashi_2014,Sakai_2014}. This is in contrast to most known young Class 0/I sources, where observations of a prominent outflow have become ubiquitous. Additionally, most of the emission detected in H$_2$CO, the only other molecule besides the CO isotopologues detected around IRS5N, is at a tentative level of 3$\sigma - 5\sigma$.

The curious case of low emission around IRS5N has also been noted by previous studies \citep{Nutter_2005,Lindberg_2014}. \citet{Lindberg_2014} specifically noted that only marginal residuals remained in the \textit{Herschel}/PACS maps of the region when assuming that all emission originated from the IRS5 source. Recent studies suggest that previously thought young Class 0 objects exhibiting weak molecular line emission and lack prominent high-velocity outflow structures may actually be potential candidates for first hydrostatic core (FHSC) \citep{Busch_2020,Maureira_2020,Dutta_2022}. These FHSC objects, however, have a relatively short lifetime of $\sim$10$^3$ yr and simulations predict their luminosities to be $\sim$0.1 $L_\odot$ with the mass of the central source of $\lesssim$0.1 $M_\odot$ \citep{Commercon_2012,Tomida_2015,Maureira_2020}. Considering that IRS5N has a bolometric luminosity of 1.40 $L_\odot$ and a protostellar mass of 0.3 $M_\odot$, it has already progressed well beyond the FHSC stage and this most likely is not the explanation for the observed low emission and lack of outflow. Nonetheless, given the presence of a massive envelope of $\sim$1.2 $M_\odot$ surrounding IRS5N, it is likely to become much more massive in the future.

The peculiarity of the molecular emission characteristics of IRS5N are most likely explained by the complexity of the Coronet region. IRS7B, another YSO source of eDisk from the Coronet region, also seems to lack an outflow in the spectral lines \citep{Ohashi_2023}. The Coronet hosts numerous YSOs and Molecular Hydrogen emission-line Objects (MHOs) with more than 20 Herbig-Haro (HH) objects \citep[see][and references therein]{Wang_2004}. Such an environment might be affecting the molecular emission seen from these sources. $^{12}$CO, being optically thick, is the most affected. We do observe \thrco~emission in the North-South direction of the source, roughly in the direction where the outflow is expected. However, this is only seen at low velocity, red-shifted channel maps. \C18O appears to be the least affected among the CO isotopologues as it is the most optically thin of the three and is not as hidden behind the optically thick emission from the cloud like \tlvco~and \thrco~making it mostly sensitive to the inner disk where the CO is evaporated from the dust grains \citep{Jorgensen_2015}.


\section{Conclusions}\label{sec:conclusion}
We have presented high-resolution, high-sensitivity observations of the protostar IRS5N and its surroundings as part of the eDisk ALMA Large program. Our ALMA band 6 observation had a continuum angular resolution of $\sim0\farcs05$ ($\sim$8 au) and molecular line emission from \C18O, \tlvco, \thrco, and H$_2$CO. The main results of the paper are as follows:

\begin{enumerate}

    \item The 1.3 mm dust continuum emission traces protostellar disks around IRS5N and IRS5. The continuum emissions appear smooth, with no apparent substructures in either source. However, the disk of IRS5N shows brightness asymmetry in the minor axis, with the southern region appearing brighter than the northern region. The asymmetry can be attributed to the geometrical effects of optically thick emission and flaring of the disk.
    
    \item IRS5N has a disk radius of $\sim$62 au elongated along the northeast to southwest direction with a P.A. of 81.10$^{\circ}$. IRS5a has a much smaller disk radius of $\sim$13 au with a P.A. of $\sim$85$^{\circ}$. The disk of IRS5b remains unresolved. Using the total integrated intensity of each source and assuming a temperature of $T=20$ K, which is a typical dust temperature for Class II disks, the estimated disk masses for IRS5N, IRS5a, and IRS5b are 0.02, 9.18 $\times$ 10$^{-4}$, and 6.48 $\times$ 10$^{-5}$ $M_{\odot}$, respectively. At a temperature of $T=47$ K based on radiative transfer, the estimated disk masses for IRS5N and IRS5a are 6.65 $\times$ 10$^{-3}$ and 3.20 $\times$ 10$^{-4}$ $M_{\odot}$, respectively.
    
    \item Disk rotation is observed in the \C18O~emission around IRS5N, with the blue- and red-shifted emission separated along the major axis of the disk. PV analysis of the emission reveals the disk is in Keplerian rotation. The stellar mass of the central source of IRS5N is estimated to be $\sim$0.3 $M_{\odot}$. 

    \item Using a 1D dust radiative transfer model, the estimated envelope mass around IRS5N is 1.2 $M_{\odot}$. The envelope mass is much greater than the disk mass of 0.02 $M_{\odot}$ and stellar mass of 0.3 $M_{\odot}$ indicating IRS5N is a highly embedded protostar.
    
    \item The \tlvco~and \thrco~maps towards IRS5N are complex and lack any apparent indication of an outflow or cavity. In contrast, the \tlvco~maps around IRS5 show emission streaming from IRS5b to IRS5a, tracing the gas connecting to the disk-like structure around the latter. This observation potentially suggests material transport between the two sources.


\end{enumerate}

\section*{Acknowledgments}
This paper makes use of the following ALMA data: ADS/ JAO.ALMA\#2019.1.00261.L. ALMA is a partnership of ESO (representing its member states), NSF (USA), and NINS (Japan), together with NRC (Canada), MOST and ASIAA (Taiwan), and KASI (Republic of Korea), in cooperation with the Republic of Chile. The Joint ALMA Observatory is operated by ESO, AUI/NRAO, and NAOJ. 
The National Radio Astronomy Observatory is a facility of the National Science Foundation operated under cooperative agreement by Associated Universities, Inc. 
R.S, J.K.J, and S.G. acknowledge support from the Independent Research Fund Denmark (grant No. 0135-00123B). 
S.T. is supported by JSPS KAKENHI grant Nos. 21H00048 and 21H04495. This work was supported by NAOJ ALMA Scientific Research Grant Code 2022-20A. 
L.W.L. acknowledges support from NSF AST-2108794. J.J.T. acknowledges support from NASA XRP 80NSSC 22K1159. 
N.O. acknowledges support from National Science and Technology Council (NSTC) in Taiwan through grants NSTC 109-2112-M-001-051 and 110-2112-M-001-031. 
S.P.L. and T.J.T. acknowledge grants from the National Science and Technology Council of Taiwan 106-2119-M-007-021-MY3 and 109-2112-M-007-010-MY3. 
I.d.G. acknowledges support from grant PID2020-114461GB-I00, funded by MCIN/AEI/ 10.13039/501100011033. 
Z.Y.D.L. acknowledges support from NASA 80NSSCK1095, the Jefferson Scholars Founda- tion, the NRAO ALMA Student Observing Support (SOS) SOSPA8-003, the Achievements Rewards for College Scientists (ARCS) Foundation Washington Chapter, the Virginia Space Grant Consortium (VSGC), and UVA research computing (RIVANNA). 
Z.-Y.L. is supported in part by NASA NSSC20K0533 and NSF AST-1910106. W.K. was supported by the National Research Foundation of Korea (NRF) grant funded by the Korea government (MSIT; NRF-2021R1F1A1061794).
C.W.L. is supported by the Basic Science Research Program through the National Research Foundation of Korea (NRF) funded by the Ministry of Education, Science and Technology (NRF- 2019R1A2C1010851), and by the Korea Astronomy and Space Science Institute grant funded by the Korea government (MSIT; Project No. 2023-1-84000). 
H.-W.Y. acknowledges support from the National Science and Technology Council (NSTC) in Taiwan through the grant NSTC 110-2628-M-001-003- MY3 and from the Academia Sinica Career Development Award (AS-CDA-111-M03). Y.A. acknowledges support by NAOJ ALMA Scientific Research Grant code 2019-13B, Grant-in-Aid for Transformative Research Areas (A) 20H05844 and 20H05847. 
J.E.L. is supported by the National Research Foundation of Korea (NRF) grant funded by the Korean government (MSIT; grant No. 2021R1A2C1011718). J.P.W. acknowledges support from NSF AST-2107841.

\vspace{5mm}
\facilities{ALMA}


\software{CASA \citep{Mcmullin_2007}, Numpy \citep{Harris_2020}, Astropy \citep{astropy_2018}, Matplotlib \citep{Hunter_2007}, APLpy \citep{aplpy2012,aplpy2019}, SLAM (\url{https://github.com/jinshisai/SLAM})}



\restartappendixnumbering
\appendix

\section{Channel maps}\label{appendix:lines}
\begin{figure}[ht!]
  \includegraphics[width=1.0\linewidth]{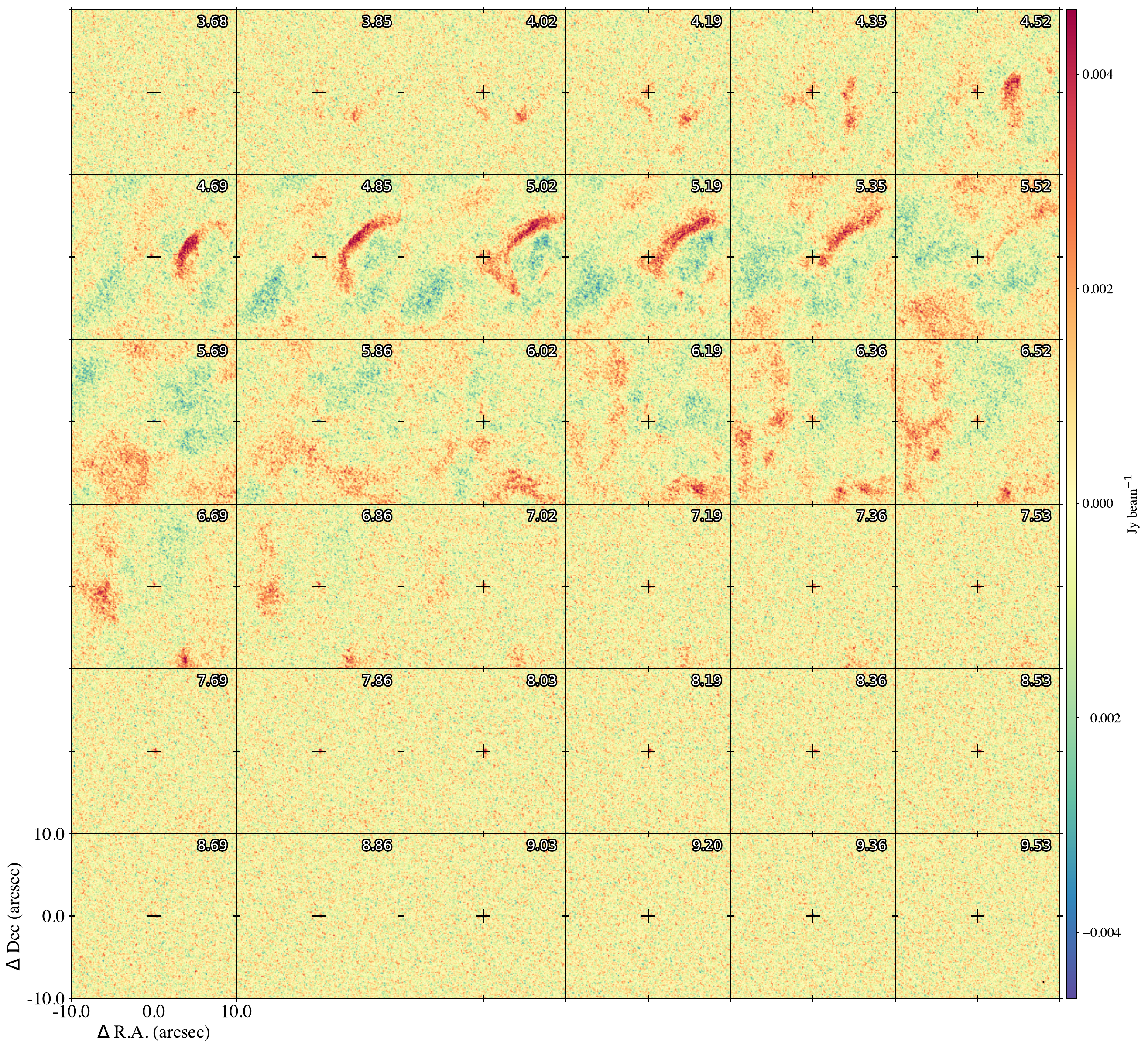} \caption{Selected channel maps showing the \C18O~(2--1) emission around IRS5N. The numbers at the top show the corresponding velocity of each channel map. The cross shows the peak position of the IRS5N continuum. Synthesized beam is shown in black on the bottom right corner of the final channel. \label{fig:appendix_c18o_channel}}
\end{figure}

\begin{figure}[ht!]
  \includegraphics[width=1.0\linewidth]{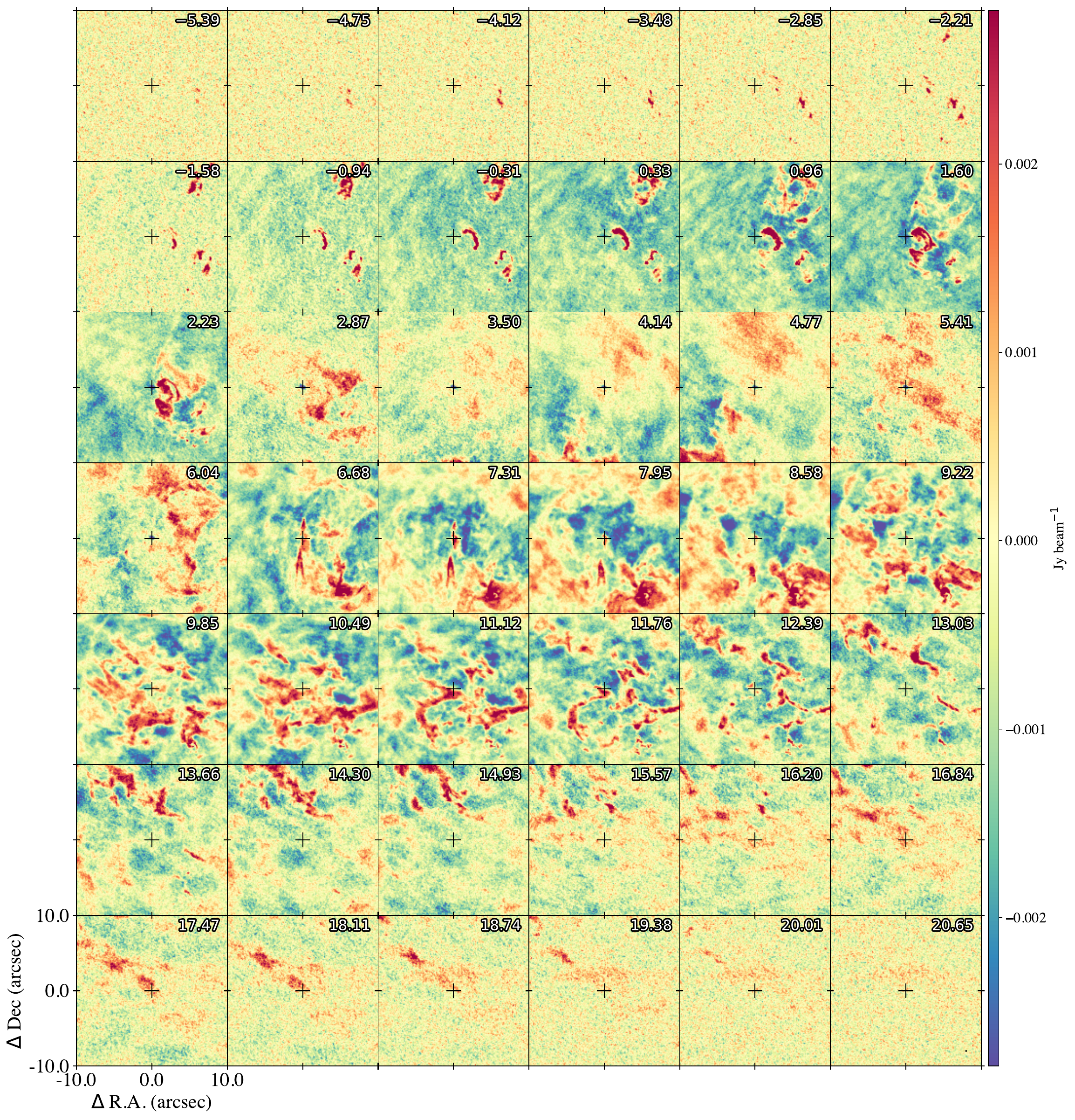} \caption{Same as Figure~\ref{fig:appendix_c18o_channel} but for \tlvco~(2--1) instead. \label{fig:appendix_12co_channel}}
\end{figure}

\begin{figure}[ht!]
  \includegraphics[width=1.0\linewidth]{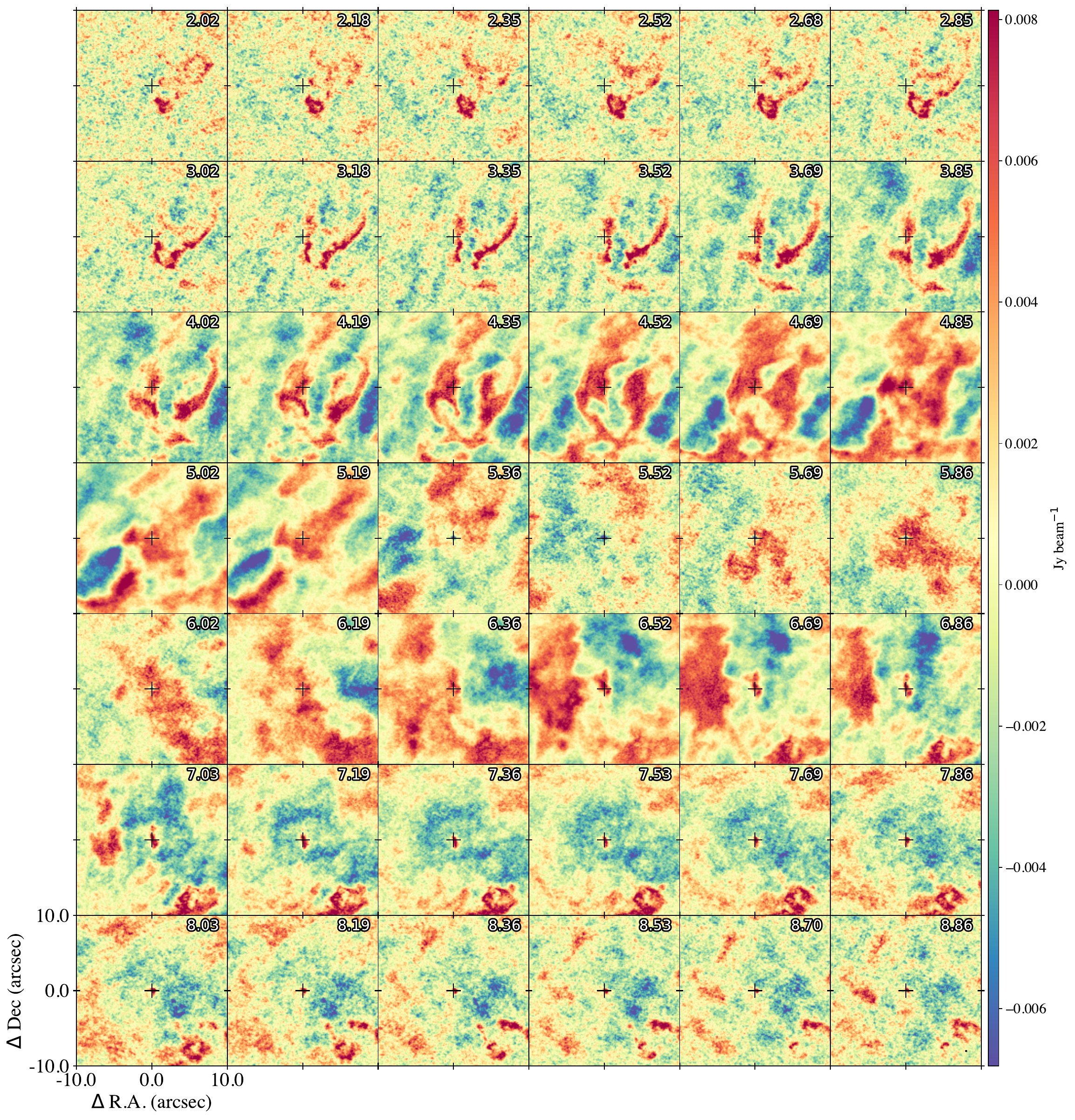} \caption{Same as Figure~\ref{fig:appendix_c18o_channel} but for \thrco~(2--1) instead. \label{fig:appendix_13co_channel}}
\end{figure}

\begin{figure}[ht!]

  \includegraphics[width=1.0\linewidth]{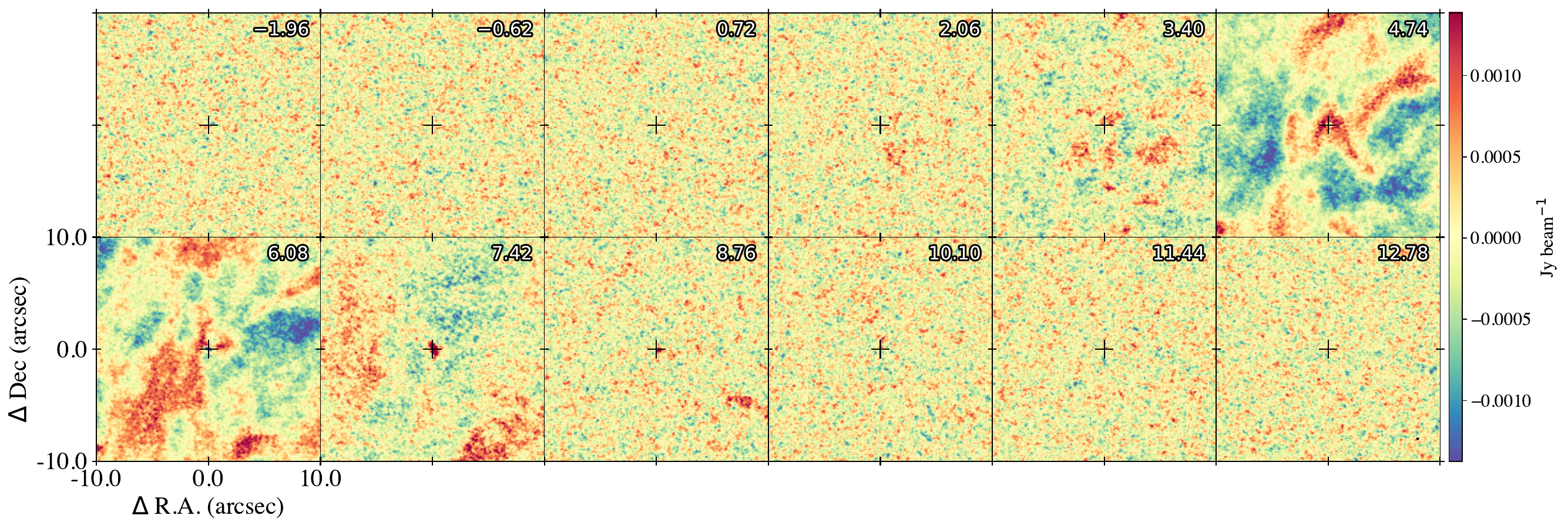} \caption{Same as Figure~\ref{fig:appendix_c18o_channel} but for H$_2$CO~(3$_{0, 3}$--2$_{0, 2}$) instead. \label{fig:appendix_h2co_3-03_2-02_channel}}
\end{figure}

\begin{figure}[ht!]
  \includegraphics[width=1.0\linewidth]{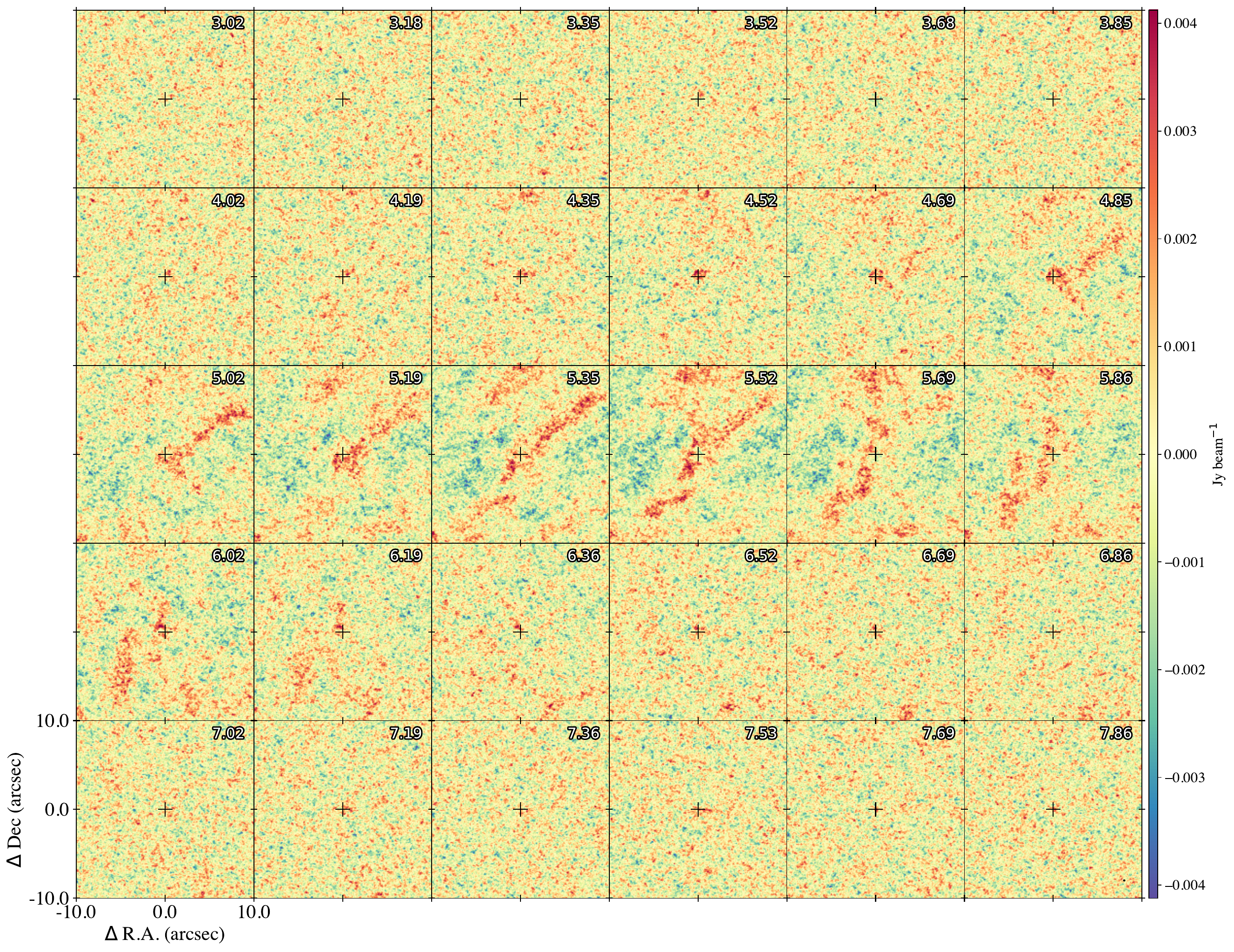} \caption{Same as Figure~\ref{fig:appendix_c18o_channel} but for H$_2$CO~(3$_{2, 1}$--2$_{2, 0}$) instead. \label{fig:appendix_h2co-3-21_2-20_channel}}
\end{figure}

\begin{figure}[ht!]
  \includegraphics[width=1.0\linewidth]{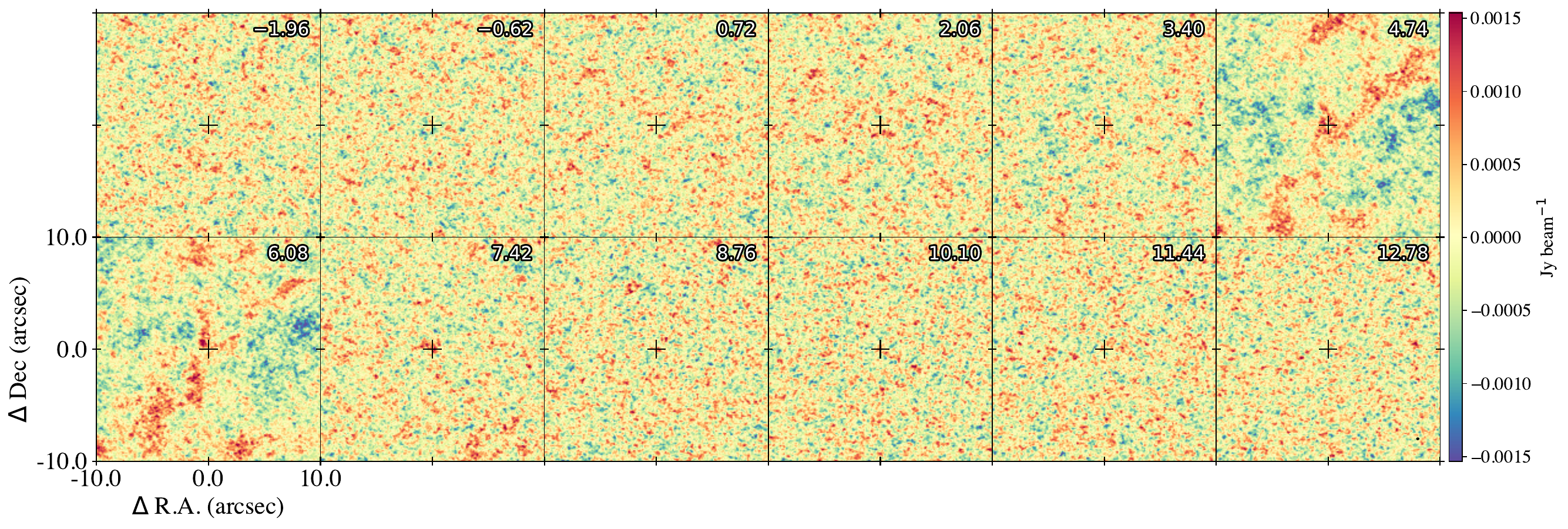} \caption{Same as Figure~\ref{fig:appendix_c18o_channel} but for H$_2$CO~(3$_{2, 2}$--2$_{2, 1}$) instead. \label{fig:appendix_h2co-3-22_2-22_channel}}
\end{figure}

\begin{figure}[ht!]
  \includegraphics[width=1.0\linewidth]{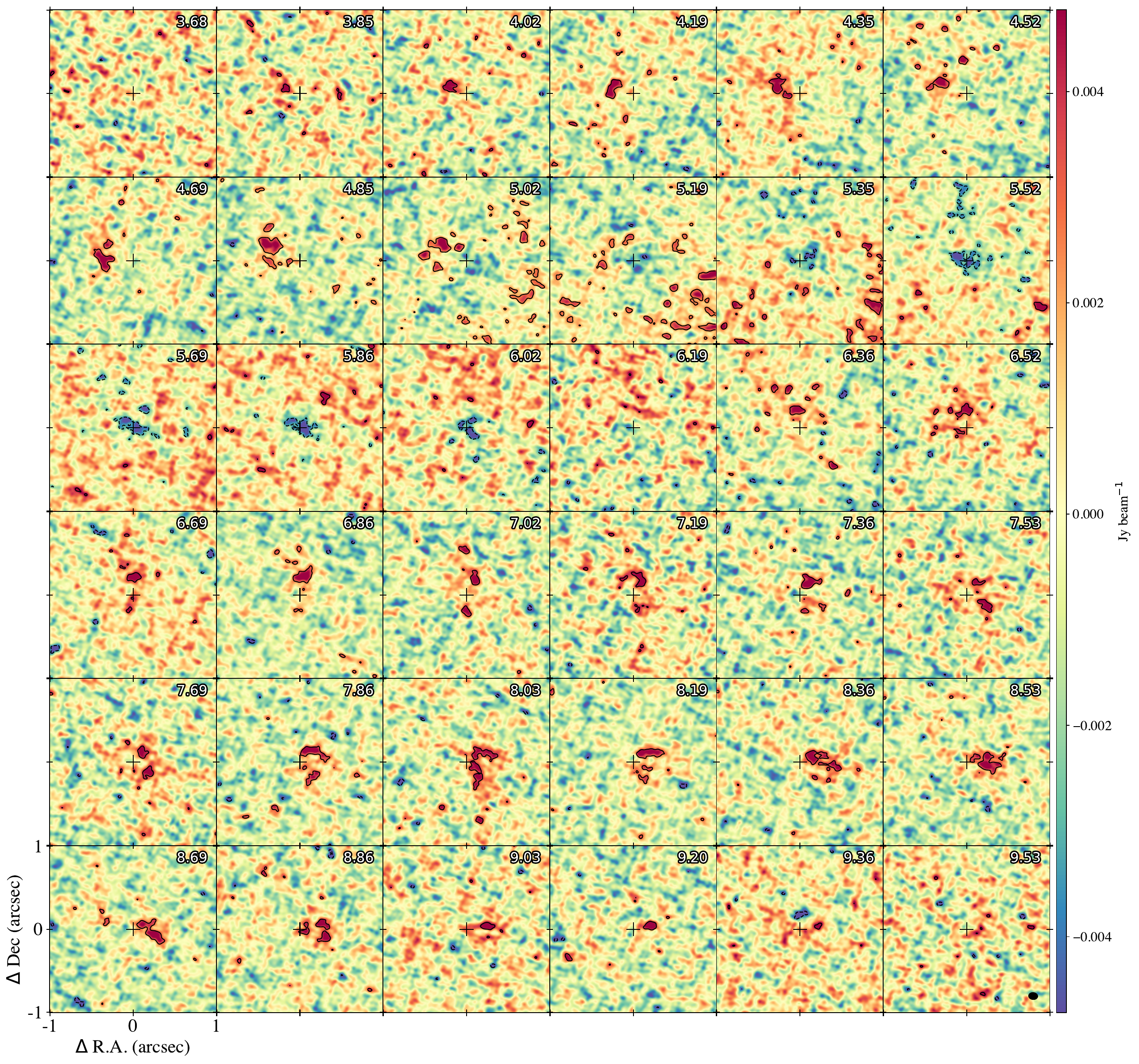} \caption{Channel maps showing the zoomed-in view of the \C18O~(2--1) emission around IRS5N. The numbers at the top show the corresponding velocity of each channel map. The solid contours represent the 3$\sigma$ level and the dashed lines show the $-3\sigma$ level. The cross shows the peak position of the IRS5N continuum. Synthesized beam is shown in black on the bottom right corner of the final channel. \label{fig:appendix_c18o_channel_zoomed}}
\end{figure}

\begin{figure}[ht!]
  \includegraphics[width=1.0\linewidth]{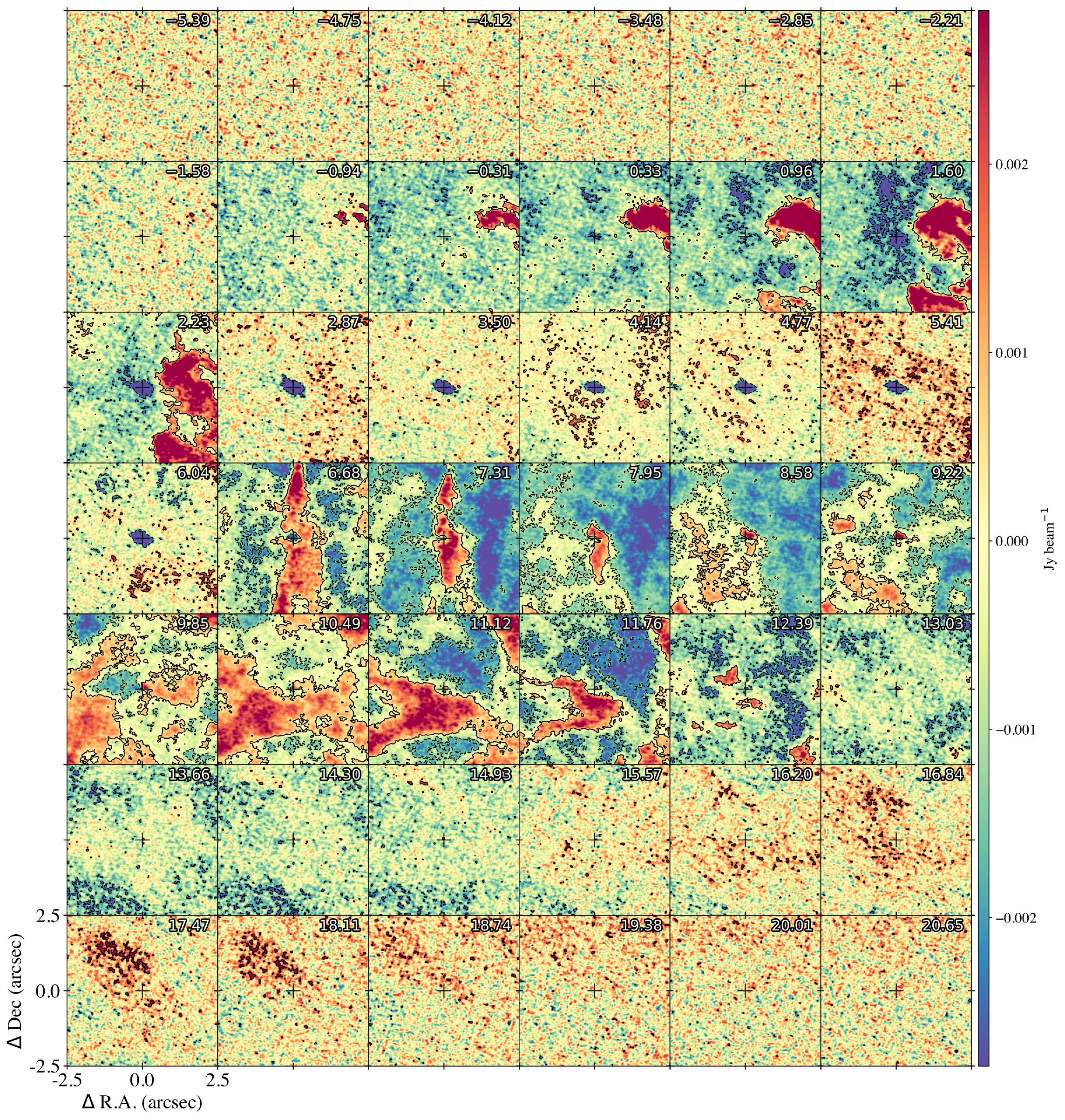} \caption{Same as Figure~\ref{fig:appendix_c18o_channel_zoomed} but for \tlvco~(2--1) instead. \label{fig:appendix_12co_channel_zoomed}}
\end{figure}

\begin{figure}[ht!]
  \includegraphics[width=1.0\linewidth]{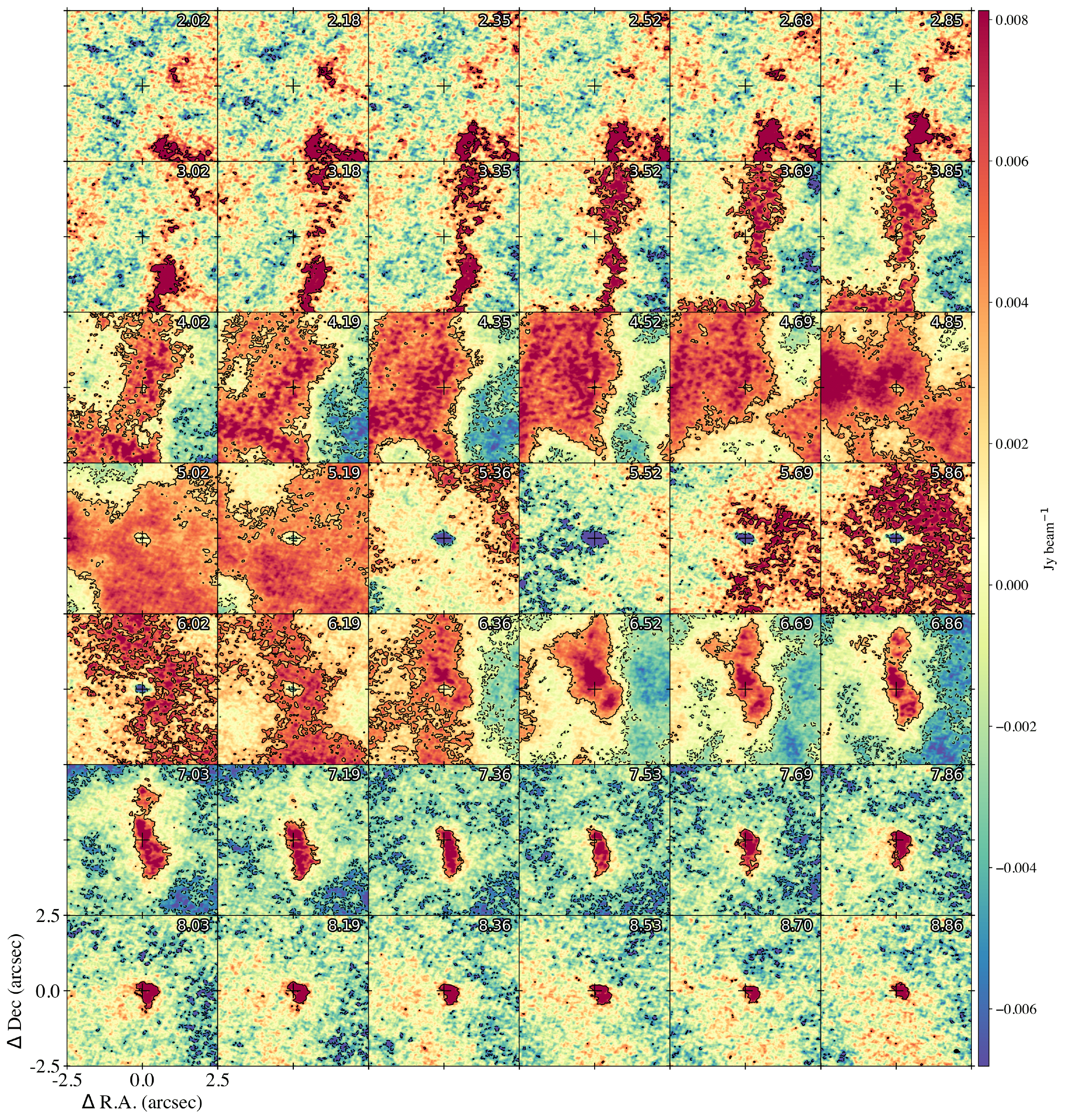} \caption{Same as Figure~\ref{fig:appendix_c18o_channel_zoomed} but for \thrco~(2--1) instead. \label{fig:appendix_13co_channel_zoomed}}
\end{figure}

\begin{figure}[ht!]

  \includegraphics[width=1.0\linewidth]{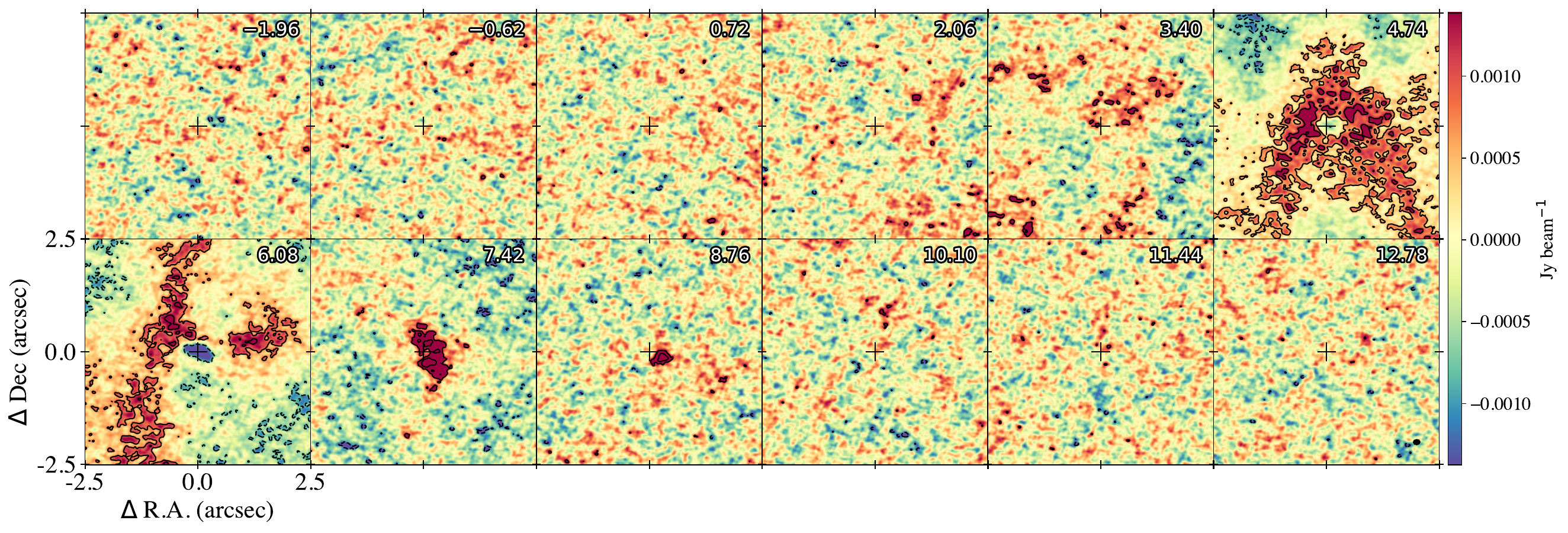} \caption{Same as Figure~\ref{fig:appendix_c18o_channel_zoomed} but for H$_2$CO~(3$_{0, 3}$--2$_{0, 2}$) instead. \label{fig:appendix_h2co_3-03_2-02_channel_zoomed}}
\end{figure}

\begin{figure}[ht!]
  \includegraphics[width=1.0\linewidth]{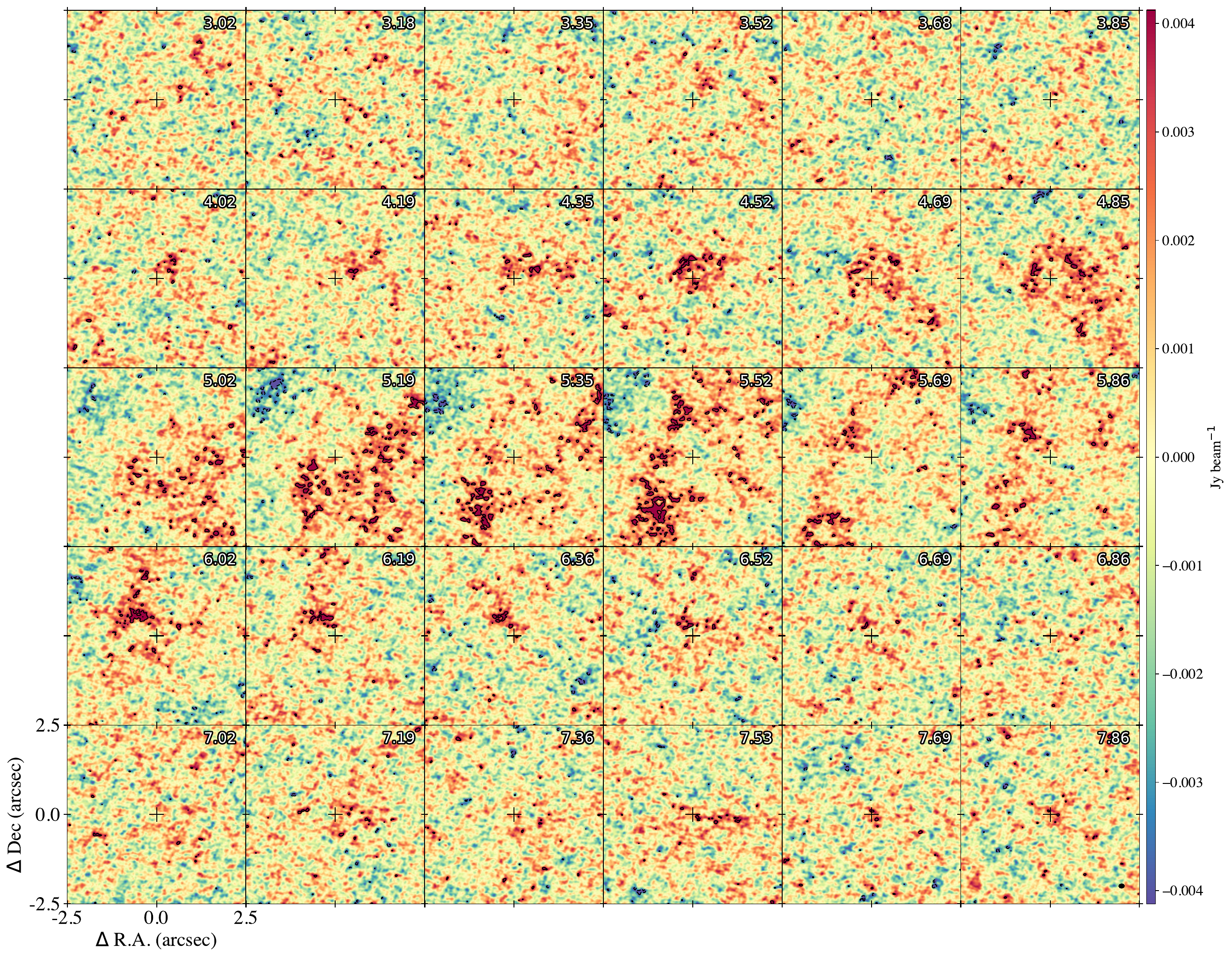} \caption{Same as Figure~\ref{fig:appendix_c18o_channel_zoomed} but for H$_2$CO~(3$_{2, 1}$--2$_{2, 0}$) instead. \label{fig:appendix_h2co-3-21_2-20_channel_zoomed}}
\end{figure}

\begin{figure}[ht!]
  \includegraphics[width=1.0\linewidth]{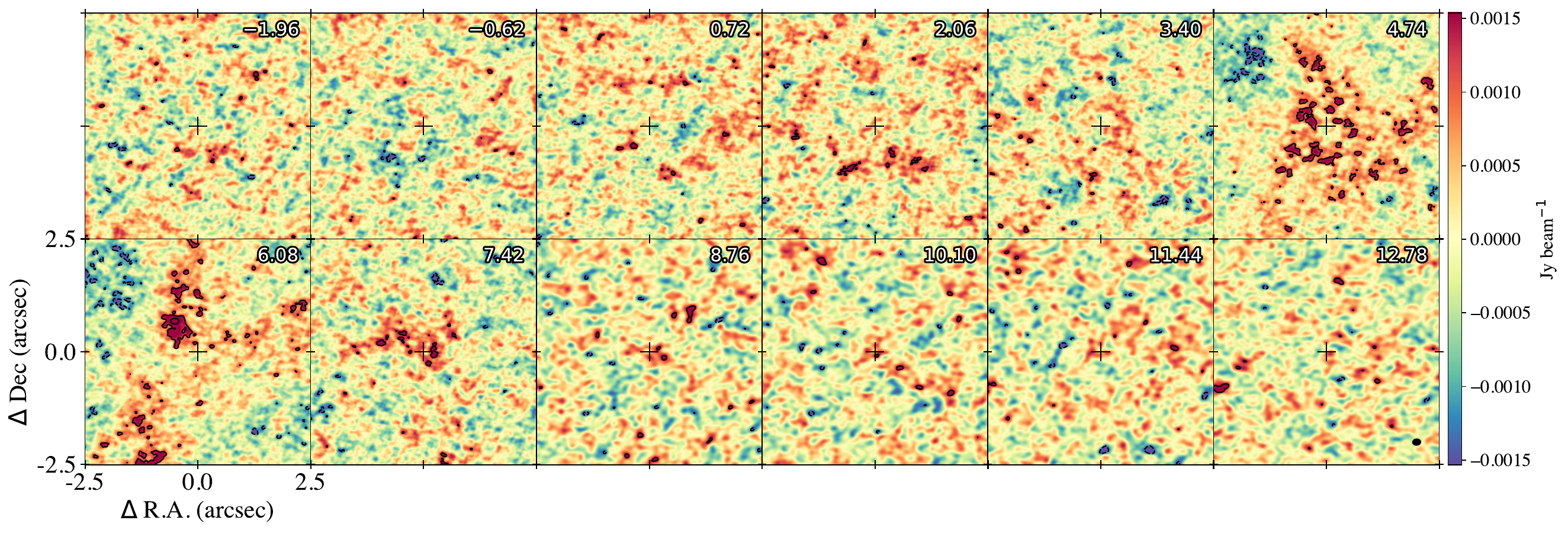} \caption{Same as Figure~\ref{fig:appendix_c18o_channel_zoomed} but for H$_2$CO~(3$_{2, 2}$--2$_{2, 1}$) instead. \label{fig:appendix_h2co-3-22_2-21_channel_zoomed}}
\end{figure}

Figure~\ref{fig:appendix_c18o_channel}--\ref{fig:appendix_h2co-3-22_2-22_channel} show the selected large-scale channel maps of $^{12}$CO, \thrco, \C18O, and H$_2$CO emission observed toward IRS5N. Figure~\ref{fig:appendix_c18o_channel_zoomed}--\ref{fig:appendix_h2co-3-22_2-21_channel_zoomed} show the zoomed-in view of the molecular emissions observed.

\bibliography{references}{}
\bibliographystyle{aasjournal}



\end{document}